\RequirePackage{silence}
\documentclass[twocolumn,twocolappendix]{aastex701}
\usepackage[utf8]{inputenc}
\usepackage[T1]{fontenc}
\let\tablenum\relax
\usepackage{siunitx}
\usepackage{amsmath}
\usepackage{booktabs}
\usepackage{comment}
\makeatletter
\def\fps@figure{!htbp}
\def\fps@table{!htbp}
\makeatother

\DeclareUnicodeCharacter{2212}{\textminus}

\providecommand{\apjl}{ApJL}
\providecommand{\apjs}{ApJS}
\providecommand{\aap}{A\&A}

\makeatletter
\@ifundefined{JournalTitle}
  {}
  {}
\makeatother

\makeatletter
\newif\ifPrismSuppressWarning
\let\PrismOrigPackageWarning\PackageWarning
\renewcommand{\PackageWarning}[2]{  \PrismSuppressWarningfalse
  \begingroup
  \edef\PrismPkgName{#1}  \edef\PrismMsg{\detokenize{#2}}  \in@{The definition of \label has changed!}{\PrismMsg}  \ifin@\global\PrismSuppressWarningtrue\fi
  \in@{Check your packages!}{\PrismMsg}  \ifin@\global\PrismSuppressWarningtrue\fi
  \ifnum\pdfstrcmp{\PrismPkgName}{hyperref}=0\relax
    \in@{Ignoring empty anchor on input line}{\PrismMsg}    \ifin@\global\PrismSuppressWarningtrue\fi
  \fi
  \endgroup
  \ifPrismSuppressWarning\else\PrismOrigPackageWarning{#1}{#2}\fi
}

\renewcommand{\PackageWarningNoLine}[2]{  \PackageWarning{#1}{#2}}
\let\PrismOrigClassWarning\ClassWarning
\renewcommand{\ClassWarning}[2]{  \PrismSuppressWarningfalse
  \begingroup
  \edef\PrismClassName{#1}  \ifnum\pdfstrcmp{\PrismClassName}{aastex701}=0\relax
    \global\PrismSuppressWarningtrue
  \fi
  \endgroup
  \ifPrismSuppressWarning\else\PrismOrigClassWarning{#1}{#2}\fi
}

\renewcommand{\ClassWarningNoLine}[2]{  \ClassWarning{#1}{#2}}
\makeatletter
\@ifundefined{Hy@Warning}{}{  \let\PrismOrigHyWarning\Hy@Warning
  \def\Hy@Warning#1{    \begingroup
    \edef\PrismMsg{\detokenize{#1}}    \in@{Ignoring empty anchor on input line}{\PrismMsg}    \ifin@\else\PrismOrigHyWarning{#1}\fi
    \endgroup
  }}
\let\PrismOrigLatexWarning\@latex@warning
\def\@latex@warning#1{  \PrismSuppressWarningfalse
  \begingroup
  \edef\PrismMsg{\detokenize{#1}}  \edef\PrismFirstPageWarn{\detokenize{Label `FirstPage' multiply defined.}}  \edef\PrismMultiWarn{\detokenize{There were multiply-defined labels.}}  \ifnum\pdfstrcmp{\PrismMsg}{\PrismFirstPageWarn}=0\relax\global\PrismSuppressWarningtrue\fi
  \ifnum\pdfstrcmp{\PrismMsg}{\PrismMultiWarn}=0\relax\global\PrismSuppressWarningtrue\fi
  \endgroup
  \ifPrismSuppressWarning\else\PrismOrigLatexWarning{#1}\fi
}
\makeatother

\newcommand{\muas}{\,$\mu$as}
\newcommand{\valmuas}[1]{\ensuremath{#1\,\mu\mathrm{as}}}

\hypersetup{bookmarks=false}

\begin{document}

\title{Observational Evidence of an Internal Shock Collision in the Relativistic Jet of the Nearby Radio Galaxy 3C 111}
\shorttitle{Internal Shock Collision in 3C 111}
\shortauthors{Kawamura et al.}
\journalinfo{Accepted for publication}

\newcommand{\AffUEC}{Graduate School of Informatics and Engineering, The University of Electro-Communications, Chofu, Tokyo 182-8585, Japan}
\newcommand{\AffNAOJMitaka}{National Astronomical Observatory of Japan, 2-21-1 Osawa, Mitaka, Tokyo 181-8588, Japan}
\newcommand{\AffKogakuin}{Kogakuin University of Technology \& Engineering, Academic Support Center, 2665-1 Nakano, Hachioji, Tokyo 192-0015, Japan}
\newcommand{\AffUTokyo}{Department of Astronomy, Graduate School of Science, The University of Tokyo, 7-3-1 Hongo, Bunkyo-ku, Tokyo 113-0033, Japan}
\newcommand{\AffNAOJMizusawa}{Mizusawa VLBI Observatory, National Astronomical Observatory of Japan, 2-12 Mizusawa Hoshigaoka-cho, Oshu-shi, Iwate 023-0801, Japan}
\newcommand{\AffTokyoElectron}{Tokyo Electron Technology Solutions Limited, Iwate 023-1101, Japan}
\newcommand{\AffNagoyaCity}{Graduate School of Science, Nagoya City University, Yamanohata 1, Mizuho-cho, Mizuho-ku, Nagoya 467-8501, Aichi, Japan}
\newcommand{\AffHeriotWatt}{Institute of Sensors, Signals and Systems, Heriot-Watt University, Edinburgh EH14 4AS, UK}
\newcommand{\AffHaystack}{Massachusetts Institute of Technology Haystack Observatory, 99 Millstone Rd, Westford, MA 01880, USA}
\newcommand{\AffCfA}{Center for Astrophysics | Harvard \& Smithsonian, 60 Garden Street, Cambridge, MA 02138, USA}
\newcommand{\AffINAFIRA}{Istituto Nazionale di Astrofisica, Istituto di Radioastronomia, Via P. Gobetti, 101, 40129 Bologna, Italy}
\newcommand{\AffUnibo}{Dipartimento di Fisica e Astronomia, Bologna University, 40129 Bologna, Italy}
\newcommand{\AffXAO}{Xinjiang Astronomical Observatory, Chinese Academy of Sciences, 150 Science 1-Street, Urumqi 830011, People's Republic of China}
\newcommand{\AffMPIfR}{Max-Planck-Institut für Radioastronomie, Auf dem Hügel 69, 53121 Bonn, Germany}
\newcommand{\AffIAA}{Instituto de Astrofísica de Andalucía-CSIC, Glorieta de la Astronomía s/n, 18008 Granada, Spain}
\newcommand{\AffINAFCagliari}{Istituto Nazionale di Astrofisica---Osservatorio Astronomico di Cagliari, Via della Scienza 5, 09047 Selargius, Italy}
\newcommand{\AffYonsei}{Department of Astronomy, Yonsei University, Yonsei-ro 50, Seodaemun-gu, Seoul 03722, Republic of Korea}
\newcommand{\AffKASI}{Korea Astronomy and Space Science Institute, Daedeok-daero 776, Yuseong-gu, Daejeon 34055, Republic of Korea}
\newcommand{\AffUST}{Department of Astronomy and Space Science, University of Science and Technology, Gajeong-ro 217, Yuseong-gu, Daejeon 34113, Republic of Korea}
\newcommand{\AffBHI}{Black Hole Initiative at Harvard University, 20 Garden Street, Cambridge, MA 02138, USA}
\newcommand{\AffINAFCatania}{Istituto Nazionale di Astrofisica--Osservatorio Astrofisico di Catania, Via Santa Sofia 78, I-95123 Catania, Italy}
\newcommand{\AffIAAStP}{Institute of Applied Astronomy, Russian Academy of Sciences, Kutuzova Embankment 10, 191187 St. Petersburg, Russia}

\author{Kenzo Kawamura}
\email[show]{kkawamura.astro@gmail.com}
\affiliation{\AffUEC}
\affiliation{\AffNAOJMitaka}

\author{Motoki Kino}
\email{motoki.kino@gmail.com}
\affiliation{\AffNAOJMitaka}
\affiliation{\AffKogakuin}

\author{Mareki Honma}
\email{mareki.honma@nao.ac.jp}
\affiliation{\AffNAOJMitaka}
\affiliation{\AffNAOJMizusawa}

\author{Kazuhiro Hada}
\email{hada@nsc.nagoya-cu.ac.jp}
\affiliation{\AffNAOJMizusawa}
\affiliation{\AffNagoyaCity}

\author{Paul Tiede}
\email{paul.tiede@fas.harvard.edu}
\affiliation{\AffCfA}
\affiliation{\AffBHI}

\author[0000-0003-4540-4095]{Evgeniya Kravchenko}
\email{evgenia.v.kravchenko@gmail.com}
\affiliation{Independent researcher}

\author{Marcello Giroletti}
\email{marcello.giroletti@inaf.it}
\affiliation{\AffINAFIRA}

\author{Rocco Lico}
\email{rocco.lico@inaf.it}
\affiliation{\AffINAFIRA}
\affiliation{\AffIAA}

\author{Kazunori Akiyama}
\email{kazunori.akiyama@hw.ac.uk}
\affiliation{\AffHeriotWatt}
\affiliation{\AffHaystack}
\affiliation{\AffNAOJMizusawa}
\affiliation{\AffCfA}

\author{Mieko Takamura}
\email{mieko.takamura30@gmail.com}
\affiliation{\AffNAOJMizusawa}

\author{Filippo D'Ammando}
\email{dammando@ira.inaf.it}
\affiliation{\AffINAFIRA}

\author{Monica Orienti}
\email{orienti@ira.inaf.it}
\affiliation{\AffINAFIRA}

\author{Gabriele Giovannini}
\email{ggiovann@ira.inaf.it}
\affiliation{\AffINAFIRA}
\affiliation{\AffUnibo}

\author{Fumie Tazaki}
\email{fumie.tazaki@gmail.com}
\affiliation{\AffTokyoElectron}

\author{Bong Won Sohn}
\email{bwsohn@kasi.re.kr}
\affiliation{\AffYonsei}
\affiliation{\AffKASI}
\affiliation{\AffUST}

\author{Salvatore Buttaccio}
\email{salvo.buttaccio@inaf.it}
\affiliation{\AffINAFIRA}
\affiliation{\AffINAFCatania}

\author{Lang Cui}
\email{cuilang@xao.ac.cn}
\affiliation{\AffXAO}

\author{Giuseppe Maccaferri}
\email{g.maccaferri@ira.inaf.it}
\affiliation{\AffINAFIRA}

\author{Andrea Melis}
\email{andrea.melis@inaf.it}
\affiliation{\AffINAFCagliari}

\author{Alexey Melnikov}
\email{aem@iaaras.ru}
\affiliation{\AffIAAStP}

\author{Carlo Migoni}
\email{carlo.migoni@inaf.it}
\affiliation{\AffINAFCagliari}

\author{Matteo Stagni}
\email{mstagni@ira.inaf.it}
\affiliation{\AffINAFIRA}

\author{Kiyoaki Wajima}
\email{wajima@kasi.re.kr}
\affiliation{\AffKASI}
\affiliation{\AffUST}

\author{Takeshi Sakai}
\email{takeshi.sakai@uec.ac.jp}
\affiliation{\AffUEC}

\begin{abstract}
We report 10 epochs of 22 GHz East Asia To Italy: Nearly Global VLBI (EATING VLBI) observations of the nearby radio galaxy 3C~111 obtained from 2020 to 2023. Combined with VLBA data at 15 and 43 GHz, these observations yield a 64-epoch, multi-frequency sequence of high-resolution radio images.
We used high-resolution Bayesian Stokes~I imaging and Gaussian component modeling to derive light curves and knot kinematics.
We find evidence that a faster, later-emitted knot catches up with a slower, earlier knot at a deprojected distance of about 10 pc from the central engine.
The knot component responsible for this interaction propagated at relativistic speed and exhibited a rise-and-decay evolution over approximately six months.
The Stokes~I flux evolution supports the internal shock interpretation, while archival polarization images reveal a temporary increase in fractional linear polarization in the interaction region near the Stokes~I flux maximum. The timing of the 2021 GeV flare suggests that this earlier high-energy event is more naturally associated with activity near the core.
These results provide direct evidence for an internal shock in an active galactic nucleus jet and support a hybrid picture in which internal and standing shocks coexist in relativistic jets and may play different roles in their variability.
\end{abstract}


\flushbottom

\section{Introduction}
\label{sec:introduction}
About 10\% of active galactic nuclei (AGNs) are bright in the radio band and launch relativistic jets from the vicinity of the supermassive black holes (SMBHs) at their centers \citep{begelman84, urry95, lister09}.
AGNs are also the dominant extragalactic sources in high-energy $\gamma$-ray observations.
The most widely discussed scenario for such high-energy emission involves a standing shock, in which a propagating disturbance collides with a standing or quasi-stationary shock formed by jet recollimation
\footnote{{Standing, recollimation, and reconfinement shocks are sometimes discussed together with diamond-shock-like knotty structures on kpc scales. In this paper, however, we use the term recollimation shock to denote a feature set by the balance between the jet internal pressure and the external pressure. Such a recollimation feature is expected to appear only once along a jet in theoretical models with a monotonically declining external pressure profile \citep{KomissarovFalle97}.}}. 
The propagating disturbance is commonly identified with a knot component composed of non-thermal electrons.
Many $\gamma$-ray flares are attributed to such events \citep{Marscher2008, Jorstad16}, yet a subset remains unexplained.

An early theoretical model proposed that velocity irregularities within the jet can drive internal shock collisions and efficiently power radio emission \citep{rees78}. In this picture, a faster, later-emitted knot catches up with a slower, earlier-emitted knot, producing shocks that accelerate non-thermal electrons and lead to multi-wavelength brightening through inverse-Compton scattering.
In inverse-Compton scattering, low-energy seed photons are upscattered by high-energy electrons. Because the seed-photon density is generally higher closer to the SMBH, the location of the dissipation site is crucial for interpreting the resulting high-energy emission.
This framework is now widely used to explain variable high-energy emission from AGNs \citep{Spada01,Tanihata03,guetta04}, but spatially resolved observational evidence remains elusive.

In this paper, we study the broad-line radio galaxy 3C~111 (0415+379). This source is classified as an FR~II radio galaxy because of its prominent large-scale radio lobes \citep{FR1974, Hogan2011}. Its apparent jet speed reaches $v_\text{app}=8c$ \citep{Jorstad2005}. With a redshift of $z=0.049$ and an SMBH mass of $M_\text{BH}=1.8\times10^8M_\odot$ \citep[e.g.,][]{Schulz2020}, 3C~111 offers an excellent opportunity to resolve the innermost jet region. The source is bright from the radio to the $\gamma$-ray band.
In addition to launching a powerful radio jet, 3C~111 exhibits X-ray emission from its accretion disk, and flaring activity in the disk and jet appears to be correlated \citep{Chatterjee2011}.
3C~111 is also a well-known $\gamma$-ray emitter. Previous high-resolution very long baseline interferometry (VLBI) observations revealed a quasi-stationary component located about 0.3~pc from the core \citep{Jorstad17}. Such quasi-stationary components have been discussed as possible sites of $\gamma$-ray production \citep{Schulz2020}.
Moreover, 3C~111 exhibits rich polarized structures from parsec to sub-parsec scales \citep{Beuchert2018,Bartolini2025}. High fractional linear polarization and systematic EVPA structure are often regarded as signatures of shock-induced magnetic-field compression or ordering.
Therefore, 3C~111 provides a rare laboratory in which to study both high-energy emission and VLBI jet kinematics.
Finally, this source is frequently used as a calibrator, and therefore extensive archival data are available. However, no previous study has carried out quasi-simultaneous, multi-frequency monitoring of 3C~111 with a global VLBI array.

Here we present the results of a three-frequency monitoring campaign of the jet in 3C~111 from 2020 to 2023. We use 22 GHz data from the East Asia to Italy: Nearly Global VLBI array (EATING VLBI) \citep{Giovannini2023} together with 15 and 43 GHz Very Long Baseline Array (VLBA) archival data. These nearly monthly observations provide angular resolutions of better than $\valmuas{300}$, $\valmuas{475}$, and $\valmuas{160}$, respectively, and allow us to conservatively identify multiple knot components in the vicinity of the SMBH.
We also use publicly available archival CLEAN polarization images at 15~GHz as an additional diagnostic of shocks in the propagating jet.

The paper is organized as follows. In \autoref{sec:observation}, we describe the observations, and in \autoref{sec:data_reduction}, we present the calibration and analysis procedures. In \autoref{sec:result}, we present the Stokes~I imaging, archival CLEAN polarization analysis, component identification, and kinematic results. In \autoref{sec:discussion}, we discuss the relationship between the internal shock and the standing shock, the origin of the 2021 $\gamma$-ray flare, and the broader high-energy implications of the parsec-scale internal shock, before summarizing the main conclusions. Throughout this paper, we assume a flat $\Lambda$CDM cosmology \citep[e.g.,][]{PC2014} with $H_0=70\;\mathrm{km}\;\mathrm{s}^{-1}$, $\Omega_\mathrm{M}=0.3$, and $\Omega_\Lambda=0.7$. We adopt a black hole mass of $M_\text{BH}=1.8\times10^8M_\odot$ and a jet viewing angle of $\theta = (16.3 \pm 2.3)^\circ$ for 3C~111 \citep{Jorstad17}.

\section{Observations}
\label{sec:observation}

\subsection{EATING VLBI observations}

EATING VLBI is a global VLBI array that enables high-resolution monitoring observations by taking advantage of baselines up to about 10,000 km.
More recently, stations in China and Russia have joined the array, improving the coverage of intermediate-to-long baselines\citep{Giovannini2023}.
More specifically, in addition to the KaVA array, Tianma and Urumqi stations from China, Svetloe, Zelenchukskaya and Badary stations from Russia, Medicina, Noto and Sardinia stations from Italy have participated. We analyzed 10 epochs of 22 GHz monitoring observations from 2020 to 2023. The observing setup is summarized in \autoref{tab:data_table}. The data consist of 16 IFs, and each IF has a 32 MHz bandwidth. In some epochs, 14 stations participated in the observation. This provided sufficient $(u,v)$ coverage for image reconstruction. The resulting 300 \muas \; beam size and an effective Stokes~I resolution better than that of CLEAN, achieved through the Bayesian method described below enable us to identify components close to the SMBH. Therefore, the accelerating component, which is the most critical for velocity field measurement, is revealed thanks to the long East-West baseline.

\begin{table*}[t]
\centering
\caption{Observing setup of VLBI data that we used in this study.}
\label{tab:data_table}
\footnotesize
\resizebox{\textwidth}{!}{%
\begin{tabular}{ccclccc}
\hline
$\nu_{\mathrm{obs}}$ (GHz) & Program & Date & Antennas & Beam FWHM (mas) & Nominal resolution ($\mu$as) & Off-source rms (Jy pixel$^{-1}$) \\
\hline
15.18 & MOJAVE & 2020-09-03 & VLBA & $0.990\times0.580$, $177.0^\circ$ & 474 & $3.27\times10^{-6}$ \\
& & 2020-11-02 &  & $0.858\times0.593$, $177.7^\circ$ & 476 & $2.38\times10^{-6}$ \\
& & 2020-11-29 &  & $0.929\times0.613$, $177.4^\circ$ & 477 & $3.42\times10^{-6}$ \\
& & 2020-12-27 &  & $0.907\times0.607$, $178.7^\circ$ & 476 & $2.04\times10^{-6}$ \\
& & 2021-02-05 &  & $0.923\times0.578$, $3.8^\circ$ & 476 & $2.60\times10^{-6}$ \\
& & 2021-02-21 &  & $0.936\times0.603$, $1.7^\circ$ & 476 & $2.20\times10^{-6}$ \\
& & 2021-03-21 &  & $0.918\times0.604$, $176.6^\circ$ & 477 & $2.05\times10^{-6}$ \\
& & 2021-04-09 &  & $0.889\times0.603$, $179.6^\circ$ & 477 & $2.04\times10^{-6}$ \\
& & 2021-05-16 &  & $0.882\times0.675$, $11.9^\circ$ & 481 & $2.83\times10^{-6}$ \\
& & 2021-06-25 &  & $0.887\times0.605$, $1.2^\circ$ & 476 & $2.53\times10^{-6}$ \\
& & 2021-08-01 &  & $1.006\times0.602$, $179.8^\circ$ & 473 & $2.41\times10^{-6}$ \\
& & 2021-08-07 &  & $0.948\times0.617$, $1.9^\circ$ & 475 & $2.94\times10^{-6}$ \\
& & 2021-09-13 &  & $0.904\times0.603$, $174.5^\circ$ & 473 & $3.25\times10^{-6}$ \\
& & 2021-10-23 &  & $0.841\times0.577$, $170.6^\circ$ & 474 & $1.68\times10^{-5}$ \\
& & 2021-12-10 &  & $0.947\times0.621$, $5.0^\circ$ & 474 & $2.38\times10^{-6}$ \\
& & 2022-01-21 &  & $1.094\times0.588$, $176.2^\circ$ & 475 & $2.56\times10^{-6}$ \\
& & 2022-02-07 &  & $0.923\times0.634$, $2.0^\circ$ & 474 & $2.04\times10^{-6}$ \\
& & 2022-03-18 &  & $0.871\times0.640$, $174.9^\circ$ & 477 & $3.16\times10^{-6}$ \\
& & 2022-04-08 &  & $0.868\times0.638$, $171.0^\circ$ & 474 & $2.25\times10^{-6}$ \\
& & 2022-05-21 &  & $0.932\times0.632$, $0.5^\circ$ & 474 & $2.36\times10^{-6}$ \\
& & 2022-06-11 &  & $0.910\times0.598$, $178.5^\circ$ & 475 & $4.82\times10^{-6}$ \\
& & 2022-07-10 &  & $0.815\times0.581$, $0.7^\circ$ & 476 & $6.28\times10^{-6}$ \\
& & 2022-08-09 &  & $0.856\times0.576$, $173.0^\circ$ & 476 & $4.61\times10^{-6}$ \\
& & 2022-09-29 &  & $0.789\times0.525$, $171.2^\circ$ & 473 & $2.33\times10^{-5}$ \\
& & 2022-12-31 &  & $0.830\times0.603$, $0.5^\circ$ & 474 & $3.30\times10^{-6}$ \\
& & 2023-02-04 &  & $1.055\times0.634$, $176.2^\circ$ & 476 & $1.33\times10^{-6}$ \\
& & 2023-03-26 &  & $0.862\times0.622$, $171.4^\circ$ & 476 & $2.62\times10^{-6}$ \\
& & 2023-05-27 &  & $0.893\times0.553$, $178.0^\circ$ & 473 & $6.53\times10^{-6}$ \\
& & 2023-07-01 &  & $0.896\times0.634$, $173.5^\circ$ & 476 & $2.44\times10^{-6}$ \\
\hline
22.11 & EATING VLBI & 2020-09-22 & EATING - TAK & $1.025\times0.572$, $23.5^\circ$ & 298 & $8.01\times10^{-6}$ \\ 
& & 2021-02-07 & EATING - CVN & $1.144\times0.405$, $7.3^\circ$ & 286 & $1.31\times10^{-5}$ \\ 
& & 2022-01-13 & EATING - TIA, ISG & $0.864\times0.258$, $13.2^\circ$ & 296 & $1.12\times10^{-5}$ \\ 
& & 2022-03-07 & EATING - KUS & $0.999\times0.309$, $8.0^\circ$ & 296 & $1.35\times10^{-5}$ \\ 
& & 2022-05-09 & EATING - BDR, TIA & $0.616\times0.313$, $28.3^\circ$ & 324 & $1.23\times10^{-5}$ \\ 
& & 2022-10-10 & EATING - Italy & $0.784\times0.378$, $25.2^\circ$ & 347 & $1.14\times10^{-5}$ \\ 
& & 2022-12-23 & EATING - TAK, TRK & $0.650\times0.347$, $28.0^\circ$ & 298 & $1.42\times10^{-5}$ \\ 
& & 2023-02-06 & EATING - CVN, TAK, IRK & $0.799\times0.390$, $26.6^\circ$ & 292 & $2.13\times10^{-5}$ \\ 
& & 2023-04-10 & EATING - Italy, KTN, ISG & $0.664\times0.345$, $36.5^\circ$ & 353 & $2.32\times10^{-5}$ \\ 
& & 2023-06-16 & EATING - ZLC, KTN, MIZ, ISG & $0.642\times0.316$, $28.8^\circ$ & 299 & $1.99\times10^{-5}$ \\ 
\hline
42.98 & BU & 2020-08-07 & VLBA & $0.357\times0.205$, $9.2^\circ$ & 167 & $2.54\times10^{-6}$ \\ 
& & 2020-10-03 &  & $0.374\times0.191$, $4.9^\circ$ & 167 & $5.38\times10^{-6}$ \\ 
& & 2020-12-20 &  & $0.350\times0.195$, $7.9^\circ$ & 167 & $4.17\times10^{-6}$ \\ 
& & 2021-03-19 &  & $0.376\times0.199$, $178.7^\circ$ & 167 & $6.46\times10^{-6}$ \\ 
& & 2021-05-28 &  & $0.353\times0.212$, $161.1^\circ$ & 170 & $1.25\times10^{-5}$ \\ 
& & 2021-07-31 &  & $0.350\times0.186$, $6.9^\circ$ & 167 & $8.69\times10^{-6}$ \\ 
& & 2021-09-14 &  & $0.328\times0.180$, $176.2^\circ$ & 167 & $1.03\times10^{-5}$ \\ 
& & 2021-11-06 &  & $0.339\times0.201$, $176.0^\circ$ & 167 & $6.76\times10^{-6}$ \\ 
& & 2021-12-13 &  & $0.449\times0.213$, $162.1^\circ$ & 192 & $5.84\times10^{-6}$ \\ 
& & 2022-02-05 &  & $0.365\times0.184$, $175.2^\circ$ & 167 & $5.71\times10^{-6}$ \\ 
& & 2022-02-20 &  & $0.334\times0.199$, $176.5^\circ$ & 167 & $5.10\times10^{-6}$ \\ 
& & 2022-04-30 &  & $0.330\times0.221$, $14.6^\circ$ & 167 & $1.09\times10^{-5}$ \\ 
& & 2022-06-05 &  & $0.317\times0.177$, $174.5^\circ$ & 167 & $2.02\times10^{-6}$ \\ 
& & 2022-06-24 &  & $0.359\times0.190$, $150.8^\circ$ & 167 & $5.20\times10^{-6}$ \\ 
& & 2022-07-15 &  & $0.301\times0.176$, $3.7^\circ$ & 167 & $6.03\times10^{-6}$ \\ 
& & 2022-07-22 &  & $0.312\times0.178$, $3.8^\circ$ & 167 & $8.70\times10^{-6}$ \\ 
& & 2022-08-21 &  & $0.321\times0.208$, $9.1^\circ$ & 169 & $7.12\times10^{-6}$ \\ 
& & 2022-11-01 &  & $0.317\times0.192$, $4.1^\circ$ & 167 & $4.79\times10^{-6}$ \\ 
& & 2022-11-20 &  & $0.326\times0.196$, $4.0^\circ$ & 170 & $6.62\times10^{-6}$ \\ 
& & 2022-12-06 &  & $0.333\times0.172$, $166.1^\circ$ & 167 & $5.53\times10^{-6}$ \\ 
& & 2023-02-11 &  & $0.330\times0.198$, $3.2^\circ$ & 167 & $4.76\times10^{-6}$ \\ 
& & 2023-04-02 &  & $0.414\times0.244$, $21.6^\circ$ & 182 & $7.49\times10^{-6}$ \\ 
& & 2023-05-21 &  & $0.412\times0.249$, $21.5^\circ$ & 249 & $5.74\times10^{-6}$ \\ 
& & 2023-06-01 &  & $0.332\times0.208$, $15.5^\circ$ & 170 & $4.00\times10^{-6}$ \\ 
& & 2023-06-30 &  & $0.347\times0.198$, $3.1^\circ$ & 167 & $3.22\times10^{-6}$ \\ 
\hline
\end{tabular}
}
\end{table*}
However, system temperature information was missing for some stations. We had to estimate this information by other baseline amplitudes because they are necessary for amplitude calibration. This procedure corresponds to classical amplitude self-calibration, but we again utilized Bayesian estimation for this, and it provides a more statistically rigorous treatment. See \autoref{met:imaging} for more details about data reduction.

\subsection{VLBA archival data}

{
We analyzed archival VLBA data obtained quasi-simultaneously with the EATING VLBI observations. These consist of 29 epochs of 15~GHz MOJAVE visibility data \citep{Lister2018} and 25 epochs of 43~GHz BEAM-ME and BU visibility data \citep{Jorstad16} from 2020 to 2023. We applied the Stokes~I imaging and gain-estimation procedure described in \autoref{met:imaging} to these visibility data; the typical nominal resolutions adopted for the 15 and 43~GHz Stokes~I analyses are 475 and 160~$\mu$as, respectively. Separately, for the polarization analysis we directly used the publicly available, already calibrated MOJAVE CLEAN Stokes~$I$, $Q$, and $U$ FITS images at 11 epochs between 2022 April 8 and 2023 July 1. We performed no additional polarization calibration on the low-SNR cross-hand data: in particular, we did not solve for D-terms or polarization-gain corrections, and we did not transfer or reapply instrumental factors inferred in the previous Bayesian polarization analysis to the archival CLEAN products. The \verb|Comrade| imaging in this study was limited to Stokes~I.
}

\section{Data Reduction and Analysis}
\label{sec:data_reduction}

\subsection{Initial Calibration}

The 22 GHz EATING VLBI dataset consisted of raw data and lacked complete gain information. Therefore, more steps are needed for data reduction than for the other frequency dataset. First, we used \verb|AIPS|\citep{Greisen2003} for the initial calibration of the data. During fringe fitting, correlation-processing problems in the early epochs for stations outside the KaVA array caused coherence losses and reduced the SNR on intermediate-to-long baselines. As a result, we decided to flag almost all baselines that do not consist only of KaVA stations in these early epochs. However, two selected epochs nevertheless retained sufficient quality for image reconstruction. We manually flagged irregular data points using \verb|difmap|\citep{Shephered1997} after the initial calibration procedure.  

\subsection{Imaging}
\label{met:imaging}

Pixel-based Stokes I imaging and gain estimation were performed simultaneously using \verb|Comrade|\citep{Tiede2022}, which is the Bayesian modeling package of \verb|Julia| for VLBI imaging. We assumed an unresolved compact core with unknown flux density in the sky model. Also, we assigned broader probability density functions (PDFs) to stations with uncertain gains(e.g., some Russian stations in the EATING VLBI dataset) than those of stations with reliable a priori gain calibration. Since VLBI amplitude self-calibration leaves the absolute flux-density scale undetermined, we adjusted the overall image flux scale to minimize amplitude fluctuations at antennas with valid a priori gain tables (e.g., TY/GC), which serve as anchors for the amplitude calibration.

The Stokes~I gain estimation is constrained primarily by sufficiently high-SNR parallel-hand visibilities dominated by total intensity and is distinct from estimating instrumental polarization parameters from the low-SNR cross-hand data.

{
We treated the uncertainties supplied with the visibility data as random measurement errors. To account approximately for residual data-set-dependent calibration error and model mismatch, we followed the \verb|add_fractional_noise| implementation used by \verb|Comrade| and \verb|eht-imaging|/\verb|Pyehtim| \citep{Chael2018} and replaced the original thermal uncertainty $\sigma_{\mathrm{th},i}$ of each complex Stokes~I visibility $V_i$ by an effective uncertainty $\sigma_{\mathrm{eff},i}$ defined as
\begin{equation}
\sigma_{\mathrm{eff},i}^{2}
=
\sigma_{\mathrm{th},i}^{2}
+
\left(f_{\mathrm{sys}}\lvert V_i\rvert\right)^{2},
\label{eq:fractional_noise}
\end{equation}
where $f_{\mathrm{sys}}$ is an effective fractional uncertainty term. Thus, no Gaussian random deviate or other noise realization was added to $V_i$ itself; only its likelihood uncertainty was increased in quadrature.

We adopted $f_{\mathrm{sys}}=0.10$ for the 22~GHz EATING VLBI data, $0.02$ for the 15~GHz MOJAVE data, and $0.10$ for the 43~GHz BU data. The 10\% term for EATING VLBI was adopted in particular to account for the additional amplitude uncertainty associated with missing system-temperature ($T_{\mathrm{sys}}$) measurements at some stations. The approximately 10\% amplitude-calibration uncertainty reported for comparable VLBI observations \citep{Hada2012} provides only an order-of-magnitude reference; the value of $f_{\mathrm{sys}}$ used for each data set was selected empirically from the image and visibility residuals. Because the visibility data alone cannot uniquely distinguish an unmodeled instrumental contribution from source structure, we selected the fractional term manually and separately for each data set by inspecting the reconstructed images and the visibility-domain residuals. We required the off-source image artifacts to be suppressed and the root-mean-square of the normalized complex-visibility residuals, $(V_i-V_{\mathrm{model},i})/\sigma_{\mathrm{eff},i}$, to be close to unity. The fractional term is therefore an empirical error floor used to account approximately for residual calibration and model mismatch in the likelihood; it should not be interpreted as independent realizations of the absolute flux-density-scale uncertainty.
}
The prior model included only an unresolved radio core at the westernmost edge of the field of view, while extended jet emission was recovered using a highly flexible Gaussian Markov random field (GMRF) prior\citep{Tiede2025}. 
We validated the imaging results using (1) the chi-square test, (2) comparison between the model and the gain-corrected visibilities in the Fourier domain, and (3) visual inspection in the image domain. When anomalies were identified, we determined which station and epoch were responsible, flagged the affected data, and reran the imaging and gain-estimation procedure. This gain solution is used for the subsequent modeling procedure. We modeled complex visibility and closure quantities simultaneously for conservative imaging because closure quantities are independent of antenna gains\citep{Chael2018}, which are particularly uncertain for some Russian stations in the EATING VLBI array.

\subsection{Modeling}

After the imaging procedure, multi-Gaussian modeling was performed for the identification of jet components. Visibility was corrected by gain solutions that were obtained by a previous simultaneous imaging and instrumental modeling procedure (see \autoref{met:imaging}). This is analogous to the classical \verb|modelfit| procedure in \verb|difmap|, but \verb|Comrade| offers the thermal uncertainty of each parameter statistically, allowing us to evaluate the convergence of solutions safely. We adopted as many Gaussian components as could be supported with sufficiently small uncertainties, and rejected large error model for each epoch. Technically, simultaneous modeling and gain estimation is possible. However, non-pixel-based modeling cannot represent complex jet structure flexibly enough, and sometimes the gain solution converges to a clearly false value. Therefore, we did not adopt this method. We identified the sufficiently bright southernmost and westernmost component as a radio core of the jet because the position angle of the jet is about \ang{65} and has a very straightforward morphology. These core structures are bright; however, sometimes they are not the brightest component through the jet, especially at lower frequencies. However, we excluded very faint components in the southwest. They might represent counterjet structures, but we do not pursue that possibility in this paper.

For the light-curve error bars, we combined the random uncertainty of the Gaussian-component flux inferred by MCMC with a separate 10\% per-epoch flux-density-scale uncertainty. This reported-flux uncertainty is distinct from $f_{\mathrm{sys}}$: the latter is a data-set-specific error floor used to weight visibilities in the imaging and modeling likelihood, whereas the former conservatively represents the uncertainty in the final flux-density scale of each independently calibrated epoch.

\subsection{Archival CLEAN Polarization Analysis}
\label{met:clean_polarization}

{
For each of the 11 MOJAVE polarization epochs, we combined the corresponding CLEAN Stokes~$I$, $Q$, and $U$ FITS images on their native celestial-coordinate grid. We calculated the observed linear polarized intensity $P_{\mathrm{raw}}$, its Ricean-bias-corrected value $P$, the fractional linear polarization $m$, and the electric vector position angle $\chi$ as
\begin{equation}
\begin{aligned}
P_{\mathrm{raw}} &= \sqrt{Q^2+U^2}, \\
P &= \sqrt{\max\!\left(P_{\mathrm{raw}}^2-\sigma_P^2,0\right)}, \\
m &= \frac{P}{I}, \\
\chi &= \frac{1}{2}\operatorname{atan2}(U,Q).
\end{aligned}
\label{eq:clean_polarization}
\end{equation}
Here, $\sigma_I$ and $\sigma_P$ are the Stokes~I and linear-polarization rms noise levels published for each MOJAVE epoch. For display in the CLEAN-map montage, the polarized-intensity color and EVPA ticks are shown only at pixels satisfying both $I\geq5\sigma_I$ and $P_{\mathrm{raw}}\geq5\sigma_P$; the Stokes~I contours begin at $3\sigma_I$. No additional cutoff in fractional linear polarization was imposed. This display mask was not used to select pixels for the quantitative aperture measurements described below. The EVPA was calculated directly from the sky-frame Stokes~$Q$ and $U$ FITS data, measured from north through east; image rotation was applied only when rendering the montage and was not applied to the quantitative EVPA values. We did not correct the EVPAs for Faraday rotation.

To locate K2 in each CLEAN image, we first convolved the full Gaussian-component Stokes~I model for that epoch with the corresponding CLEAN restoring beam. We then translated this model relative to the CLEAN Stokes~I image to maximize their normalized cross-correlation. The displacement was optimized along directions parallel and perpendicular to the jet position angle of $65^\circ$. The Gaussian-model K2 position was shifted by the resulting common displacement; when more than one Gaussian component was associated with K2, we used their flux-weighted position, and for the 2022 June 11 epoch, where K2 was not separately identified, we linearly interpolated its Cartesian sky position between the adjacent identified epochs.

We measured the mean Stokes~$I$, $Q$, and $U$ values over all pixels within a fixed circular aperture of radius 0.25~mas centered on the aligned K2 position, without applying a polarized-intensity-based pixel selection, and derived $m$ and $\chi$ using the equations above. Stokes~$I$ exceeded $5\sigma_I$ throughout each fixed K2 aperture, and the aperture-level polarized intensity was detected above $5\sigma_P$ at every epoch. We propagated the published map noise by Monte Carlo sampling. To account for positional uncertainty, we sampled the Gaussian-component position errors and estimated the alignment uncertainty by residual bootstrap: the beam-correlated residual between the CLEAN Stokes~I image and the aligned Gaussian model was circularly displaced and the alignment was refitted. The central 16th--84th percentile interval gives the quoted 68\% uncertainty. We report both the map-noise-only interval and the total interval including the K2-position and alignment uncertainties.
}

\subsection{Calculation and Evaluation of the Lorentz Factor and Error}

We calculated the intrinsic velocity of the components $\beta$ by \autoref{eq:beta_from_betaapp},

\begin{equation}
  \beta = \frac{\beta_{\rm app}}{\beta_{\rm app}\cos\theta + \sin\theta}.
  \label{eq:beta_from_betaapp}
\end{equation}
where $\beta_{\rm app}$ is the apparent velocity of the components and $\theta$ is the viewing angle.
The apparent velocity, $\beta_{\rm app}$, and its uncertainty, $\sigma_{\beta_{\rm app}}$, are obtained by performing a linear regression on each data point of the knot.

Finally, we obtain the Lorentz Factor $\Gamma$ by \autoref{eq:gamma_def},

\begin{equation}
  \Gamma = \frac{1}{\sqrt{1-\beta^2}}.
  \label{eq:gamma_def}
\end{equation}

We propagate both $\sigma_{\beta_\text{app}}$ and $\sigma_\theta$ to the Lorentz factor
$\Gamma(\beta_\text{app},\theta)$ by \autoref{eq:sigma_gamma_mu_theta} using a first-order approximation and assuming that the two
sources of uncertainty are independent
\begin{equation}
  \sigma_\Gamma^2 \simeq
  \left(\frac{\partial \Gamma}{\partial \beta_\text{app}}\right)^2 \sigma_{\beta_\text{app}}^2
  +
  \left(\frac{\partial \Gamma}{\partial \theta}\right)^2 \sigma_\theta^2,
  \label{eq:sigma_gamma_mu_theta}
\end{equation}
where the partial derivatives are evaluated at $(\beta_\text{app},\theta_0)$.

\subsection{Hough Transform}
\label{met:Hough_transform}

The Hough transform is the mathematical method that is used not only for image processing but also for the detection of a straight line in discrete data points in two-dimensional parameter space\citep{Duda+1972}.
{ 3C 111 shows a well-collimated, single-ridge-like, straight jet morphology from 2020 to 2023. We obtained a position angle of $66.15\pm4.19^\circ$ for the 3C 111 components. The aspect ratio of the fluctuations perpendicular and parallel to the jet axis is $\tan(4.19^\circ)\simeq0.07$. Therefore, we treat the core separation as a one-dimensional radial distance and do not further investigate perpendicular fluctuations. Therefore, 
}
we utilized this method to detect straight lines within the multi-Gaussian components, which are described by radial distance and time. 
Also, we optimized the coreshift value among 3 frequencies simultaneously. 
{
To perform the Hough transform, the two parameters, time and distance, should be expressed in the same dimensionless form.
}
At first, we convert each parameter via \autoref{eq:hough_cnovert} for normalization.
{
We chose normalization factors of 1 year and 10 mas for the time and radial-distance coordinates, respectively, because the possible ranges of $t$ and $r$ should be comparable at the order-of-magnitude level, and the change in $\rho$ per unit change in $\phi$ should be sufficiently smooth over the range of $\phi$ considered.
}

\begin{equation}
\label{eq:hough_cnovert}
\begin{split}
{t_i}&=\frac{\hat{t}_i-2020.5 \;\text{year}}{{1 \;\text{year}}},\qquad\\
{r_i}&=\frac{\hat{r}_i+\Delta s_{(\nu)}}{10 \;\text{mas}}.
\end{split}
\end{equation}

where {$\hat{t}_i$ and $\hat{r}_i$} are the time and radial distance of an original data point, respectively. {$t_i$ and $r_i$} are the normalized coordinates after conversion, respectively. In addition, $\Delta s_{(\nu)}$ is the coreshift value at each frequency compared to that at 43 GHz. For each data point, we calculate the normal distance $\rho$ and the normal angle {$\phi$} that are defined in \autoref{eq:normal_parameter} while changing the value {$\phi$} within the defined range.

\begin{equation}
\label{eq:normal_parameter}
\rho = t_i\cos{\phi} + r_i\sin{\phi} .
\end{equation}

The pairs of $\rho$ and {$\phi$} vote for the accumulator defined by \autoref{eq:accumulator}.

\begin{equation}
\label{eq:accumulator}
A(\rho_k,{\phi_\ell}) \;{+}{=}\; 
\sum_{i} \mathbf{1}\!\left(\left| \rho_k - (t_i\cos{\phi_\ell} + r_i\sin{\phi_\ell}) \right| \le \tfrac{\Delta\rho}{2}\right).
\end{equation}

where $A(\rho_k,{\phi_\ell})$ is the accumulator and $(\rho_k,{\phi_\ell})$ is the pair of parameters. $\Delta\rho$ is the resolution of $\rho$. After that, we define the pairs of peak parameters using the accumulator and the peak threshold. In addition, we introduced the Nonmaximum Suppression (NMS) window\citep{Neubeck+2006}, which is expressed by \autoref{eq:peak_detection}.

\begin{equation}
\label{eq:peak_detection}
\begin{split}
A(\rho^{*},{\phi^{*}}) &\ge \tau,\\
A(\rho^{*},{\phi^{*}}) &\ge A(\rho,{\phi})
\quad \forall\,(\rho,{\phi})\in \mathcal{N}_{w_\rho,{w_\phi}}(\rho^{*},{\phi^{*}}).
\end{split}
\end{equation}

where $(\rho^{*},{\phi^{*}})$ is the candidate for the peak parameter, $\tau$ is the peak threshold, and $\mathcal{N}_{w_\rho,{w_\phi}}(\rho^{*},{\phi^{*}})$ is the NMS window. The NMS window is the neighborhood that has the parameter width $w_\rho$ and {$w_\phi$} in the $\rho$ and {$\phi$} direction centered on $(\rho^{*},{\phi^{*}})$, respectively. The parameter pairs $(\rho,{\phi})$ that have the maximum number of votes in accumulator are selected as $(\rho^{*},{\phi^{*}})$ within each NMS window so that we do not physically treat the same straight lines as different ones in our very dense data near the SMBH. 
{
All the processes described above are performed automatically. For more details, see \autoref{app:hough}.
}

We optimized the core-shift values at two frequencies, $\Delta s_{(22\;\text{GHz})}$ and $\Delta s_{(15\;\text{GHz})}$, in \autoref{eq:hough_cnovert} by selecting the combination that yielded the largest number of detected straight lines. Throughout this procedure, we adopt all parameters {$\Delta \phi = \ang{0.01}$}, $\Delta\rho = 0.004 (10\;\text{mas}+\text{year})$, $\tau=10$, $w_\rho=0.008\;(10\;\text{mas}+\text{year})$, {$w_\phi=\ang{0.2}$}. The core-shift uncertainty was uniformly set to 20~\muas, corresponding to half of the adopted Hough-transform bin width in the radial-distance direction (40~\muas).


\section{Results}
\label{sec:result}

\subsection{Imaging Overview and Collision Morphology}

\begin{figure*}[!tp]
\centering
\includegraphics[width=0.78\textwidth,height=0.72\textheight,keepaspectratio]{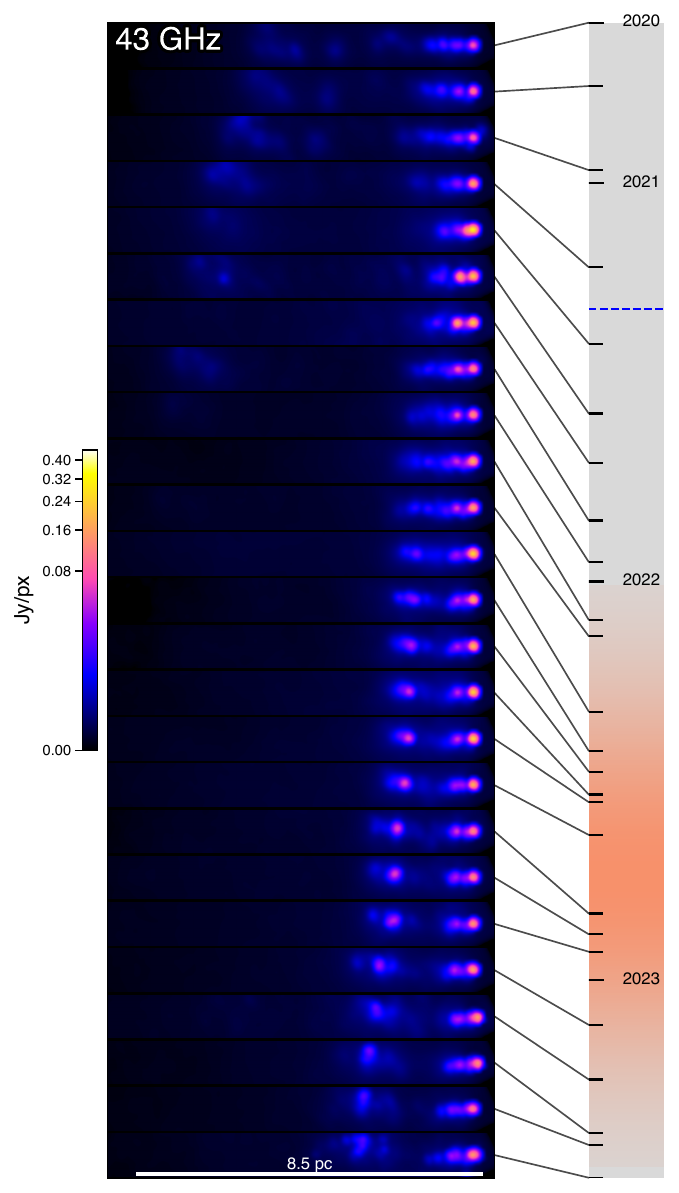}
\caption{Stokes I imaging results of the 3C 111 jet from the Boston University Blazar Monitoring Archive. The gray band on the right, connected to each image, indicates the time series of that image. The faint red band overlaid on the timeline indicates the full width at half maximum (FWHM) of the Gaussian fit to the peak of the MOJAVE light curve shown in \autoref{fig:light_curve}. The images include 25 epochs from 2020 to 2023. All images are aligned relative to the core and rotated 25 degrees counterclockwise. The radio frequency is 43 GHz and the nominal resolution is about 160 \muas. \label{fig:BU_jets}}
\end{figure*}

\begin{figure*}[!t]
\centering
\includegraphics[
  width=0.95\textwidth,
  height=0.45\textheight,
  keepaspectratio
]{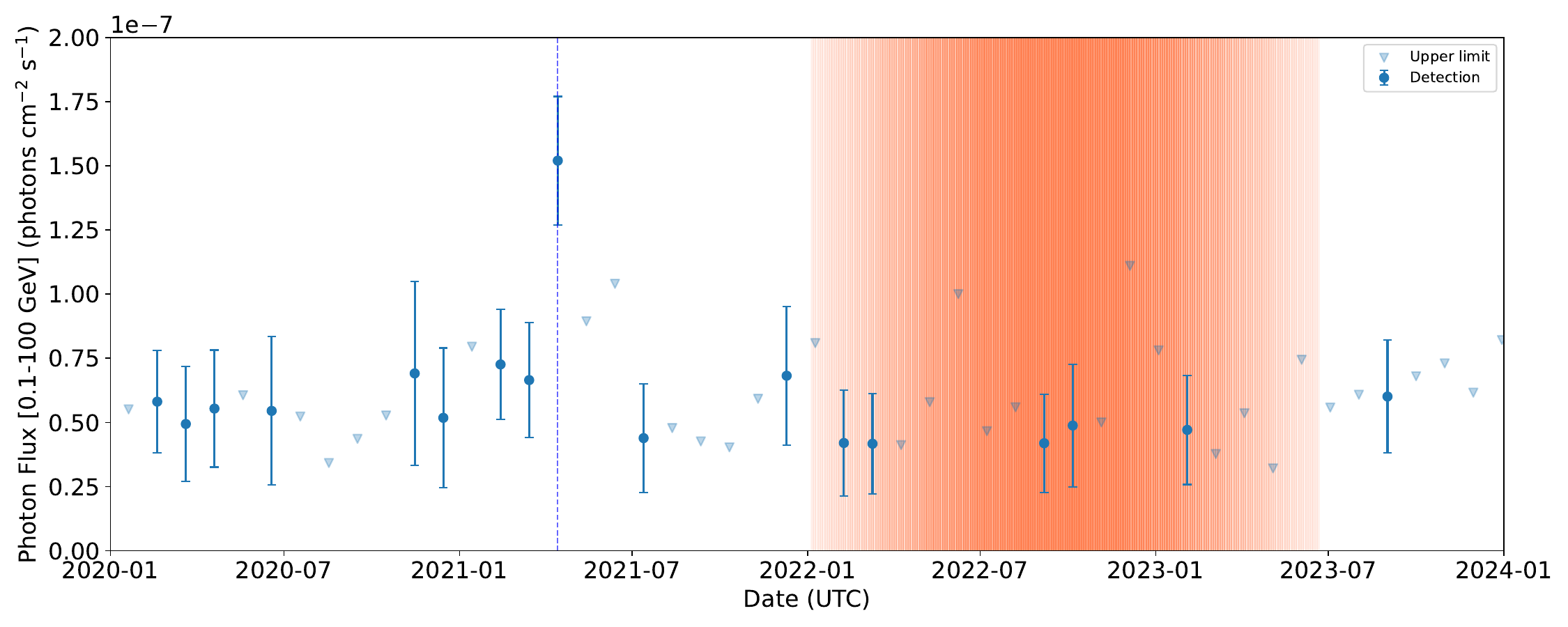}
\caption{A monthly GeV $\gamma$-ray light curve based on publicly available Fermi-LAT data from the LAT Light Curve Repository. The vertical axis shows the gamma-ray photon flux, and the horizontal axis shows time. Data points with significant detections are marked with circles, while non-significant data points are shown as downward triangles indicating upper limits. The blue vertical line marks the prominent gamma-ray burst in April 2021, and the faint red belt indicates the period during which internal shocks were identified.\label{fig:gamma_light_curve}}
\end{figure*}

\begin{figure*}[!tp]
\centering
\includegraphics[width=0.95\textwidth,height=0.82\textheight,keepaspectratio]{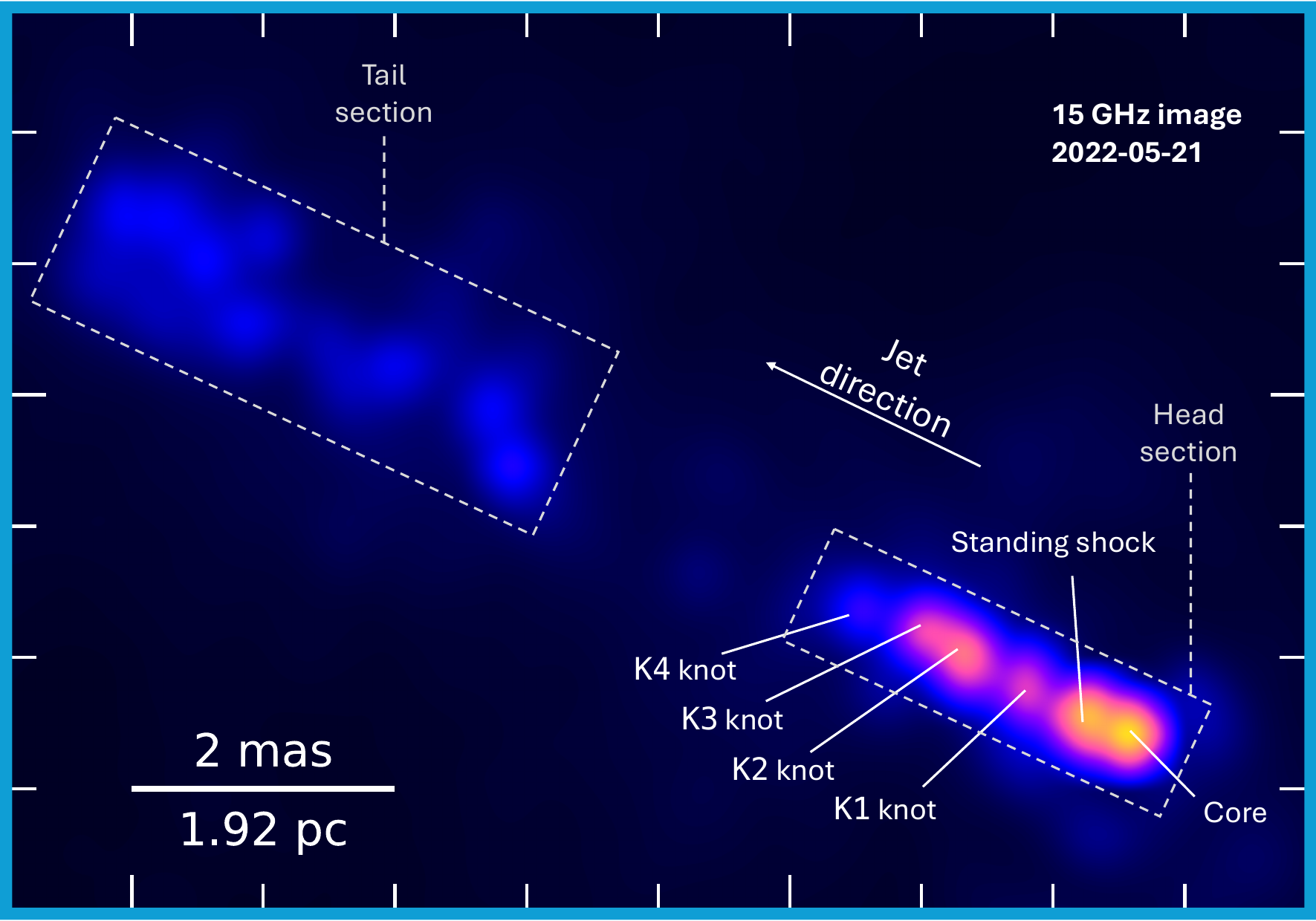}
\caption{15 GHz image of 3C~111 on May 21, 2022, in RA--DEC space. The scale bar shows 2 mas (corresponding to 1.92 pc in projected distance). The drawn arrows and text indicate the jet direction and the positions of each component. The two dotted gray rectangles mark the head and tail sections of the jet, respectively.\label{fig:jet_overview}}
\end{figure*}

\begin{figure*}[!tp]
\centering
\includegraphics[width=0.95\textwidth,height=0.85\textheight,keepaspectratio]{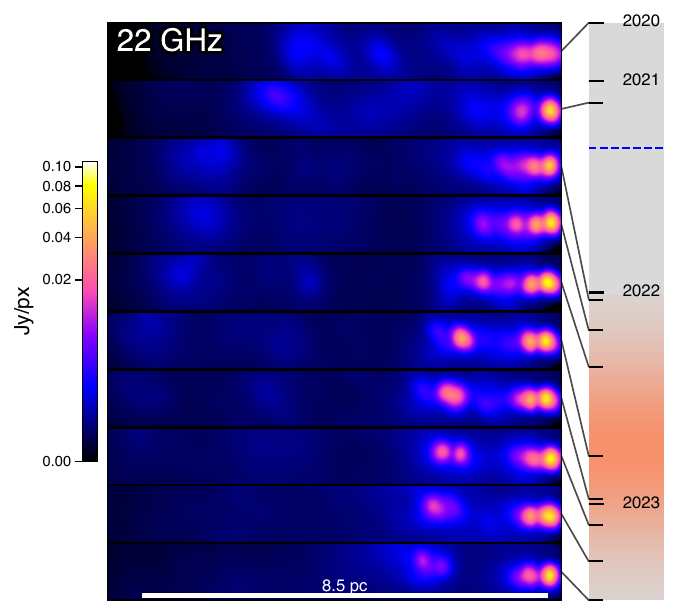}
\caption{EATING VLBI images of the 3C~111 jet (10 epochs) at 22~GHz, with a nominal resolution of $\sim 300\,\mu$as. Stokes~I (total-intensity) imaging results are shown for the 2020--2023 monitoring campaign. The gray band to the right of the panels indicates the observation time within the campaign. The faint red band overlaid on the timeline indicates the full width at half maximum (FWHM) of the Gaussian fit to the peak of the MOJAVE light curve shown in \autoref{fig:light_curve}, while the blue horizontal dotted lines mark the epochs when gamma-ray flares were detected. All images are registered to the core, rotated by $25^\circ$ counterclockwise, and aligned by shifting each map such that the westernmost pixel above a fixed intensity threshold coincides across epochs. Intensities are shown in units of Jy / pixel and displayed with a gamma stretch ($I^{0.3}$; color bar). Spatial scale bars are shown in the bottom image.\label{fig:EATING_jets}}
\end{figure*}

\begin{figure*}[!tp]
\centering
\setlength{\fboxsep}{0pt}
\includegraphics[width=0.95\textwidth,height=0.82\textheight,keepaspectratio,trim=0 0 0 2mm,clip]{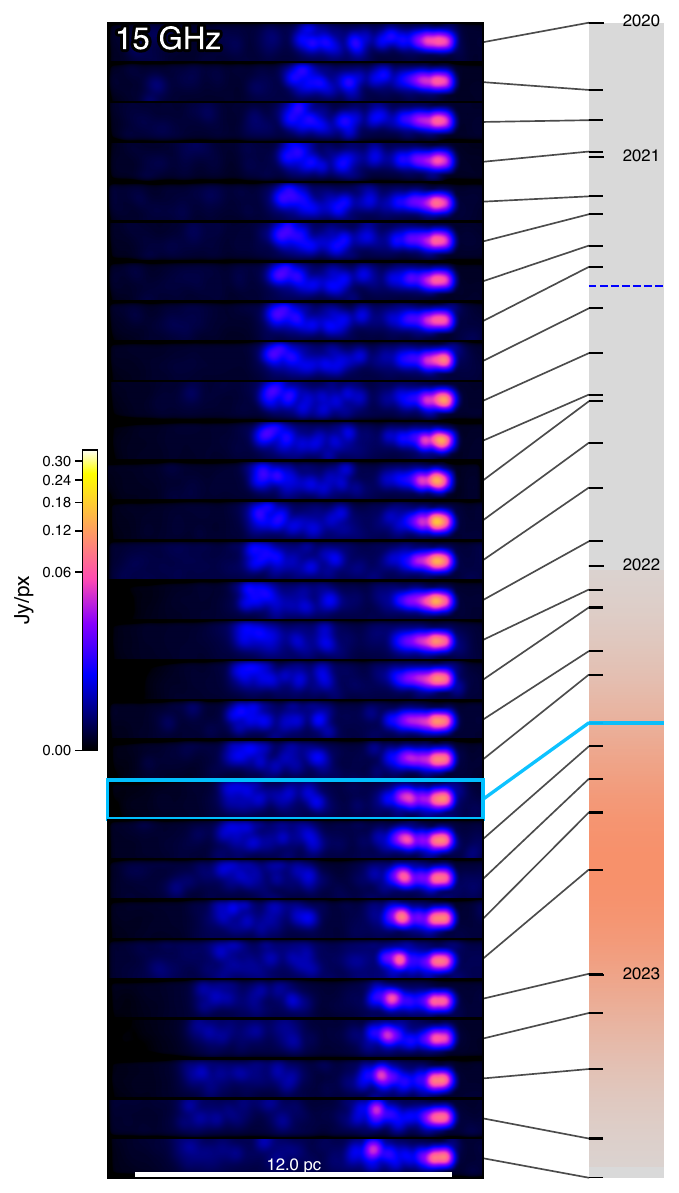}
\caption{MOJAVE archival images of the 3C~111 jet (29 epochs) at 15~GHz, with a nominal resolution of $\sim 475\,\mu$as. The epoch surrounded by the cyan frame corresponds to the 15 GHz image shown in \autoref{fig:jet_overview}. Stokes~I (total-intensity) imaging results are shown for the 2020--2023 monitoring campaign. The gray band to the right of the panels indicates the observation time within the campaign. The faint red band overlaid on the timeline indicates the full width at half maximum (FWHM) of the Gaussian fit to the peak of the MOJAVE light curve shown in \autoref{fig:light_curve}, while the blue horizontal dotted lines mark the epochs when gamma-ray flares were detected. All images are registered to the core, rotated by $25^\circ$ counterclockwise, and aligned by shifting each map such that the westernmost pixel above a fixed intensity threshold coincides across epochs. Intensities are shown in units of Jy / pixel and displayed with a gamma stretch ($I^{0.3}$; color bar). Spatial scale bars are shown in the bottom image.\label{fig:MOJAVE_jets}}
\end{figure*}

We use the 64-epoch VLBI image sequence obtained from 2020 to 2023 to trace the parsec-scale evolution of knot components in 3C~111. \autoref{fig:jet_overview}, \autoref{fig:EATING_jets}, and \autoref{fig:MOJAVE_jets} present representative views of the jet morphology and its time evolution; details of the observations and imaging procedure are given in \autoref{sec:observation} and \autoref{sec:data_reduction}.
The bright radio core lies at the rightmost edge of the images, and the jet extends eastward with a generally straight and well-collimated morphology. The component nomenclature adopted in this paper is summarized in \autoref{fig:jet_overview}.
Tracking the leading K4 knot and the trailing K2 knot through 2022, we find that their interaction begins around April 2022 and is followed by gradual brightening of K2 at a projected distance of 2--3 pc from the core. The same event is also seen in the 43 GHz BU images shown in \autoref{fig:BU_jets} \citep{Jorstad16}.
The 15 GHz images in \autoref{fig:MOJAVE_jets} consistently reveal the downstream ``tail section,'' whereas the 43 GHz images in \autoref{fig:BU_jets} resolve the upstream ``head section'' in greater detail. Across all three data sets, the jet appears segmented into ``head'' and ``tail'' regions separated by a relative emission gap.
The ``head section'' remains relatively stable until $\sim$2021.5, after which its structure changes rapidly. Immediately beforehand, the 43 GHz core brightens in coincidence with the strong GeV $\gamma$-ray flare reported on 26 April 2021 (Astronomer's Telegram 14581)(\footnote{https://www.astronomerstelegram.org/?read=14581}); the timing of this flare is indicated by the blue dotted lines in \autoref{fig:EATING_jets}, \autoref{fig:MOJAVE_jets}, and \autoref{fig:BU_jets}. See \autoref{fig:gamma_light_curve} for the corresponding Fermi-LAT $\gamma$-ray light curve\footnote{Publicly available Fermi-LAT data were obtained from the LAT Light Curve Repository: \url{https://fermi.gsfc.nasa.gov/ssc/data/access/lat/LightCurveRepository/source.html?source_name=4FGL_J0418.2+3807\#}}.

This epoch marks the onset of the subsequent dramatic evolution in the ``head'' section. 
Around July 2022, the K2 knot collides with the K4 knot and merges with it; the combined feature then propagates nearly ballistically along the downstream edge of the ``head'' section.
Its brightness peaks in late 2022 at 2--3 pc from the core and subsequently fades back to its pre-event level. These behaviors are seen at all observed frequencies and deviate from the monotonic downstream fading expected from standard radiative and adiabatic losses.
Such a collision phenomenon is consistent with the internal shock scenario.

\bigskip

\subsection{Knot Kinematics and Component Identification}

\begin{figure}[!htbp]
\centering
\includegraphics[width=0.95\linewidth]{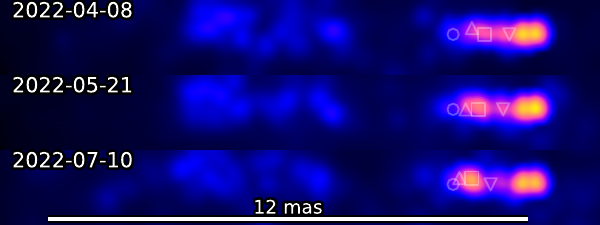}
\caption{Stokes~I images from the MOJAVE archive on 2022 April 8, May 21, and July 10, respectively. These images are also shown in \autoref{fig:MOJAVE_jets}. The white markers indicate the positions of the components highlighted in \autoref{fig:model_vs_time} and \autoref{fig:light_curve}.\label{fig:image_with_model}}
\end{figure}

\begin{figure*}[!tp]
\centering
\includegraphics[width=0.95\textwidth]{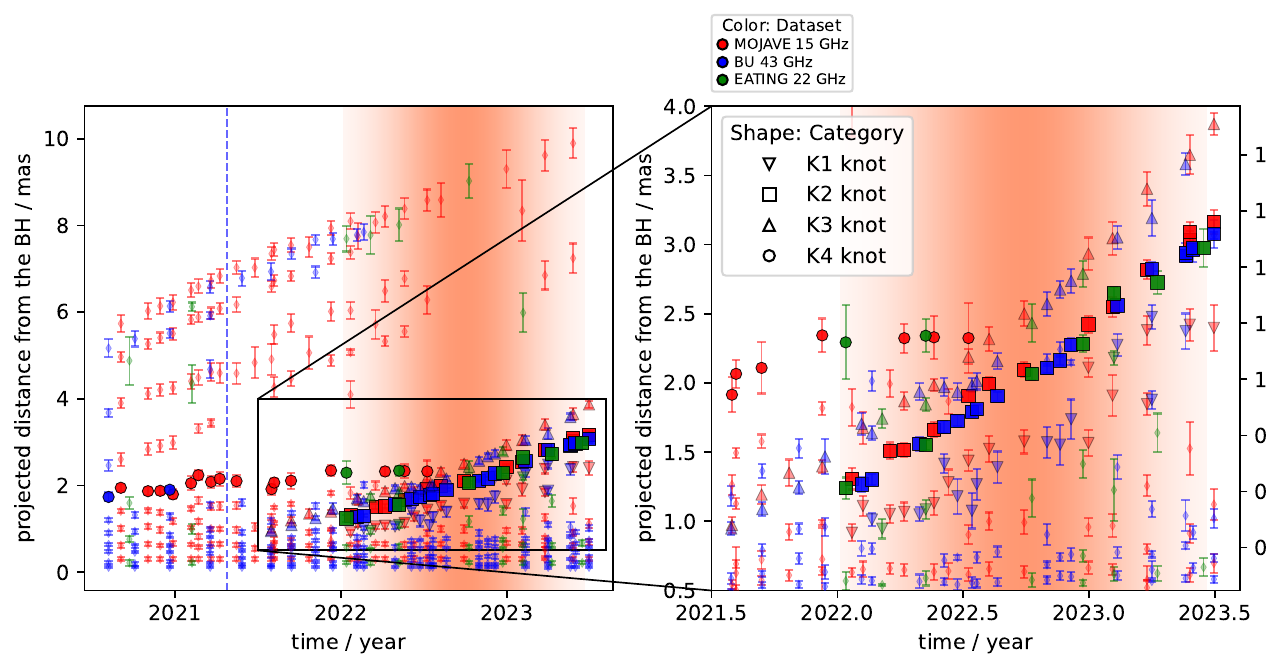}
\caption{Time evolution of the positions of the Gaussian model components. The horizontal axis shows time, and the vertical axis shows the component--black hole (BH) separation. Marker shape encodes the K1-K4 knots. Marker color indicates the dataset (MOJAVE, EATING, or BU). Error bars denote the component size (Gaussian width). The faint red band indicates the full width at half maximum (FWHM) of the Gaussian fit to the peak of the MOJAVE light curve shown in \autoref{fig:light_curve}, while the blue vertical dotted lines mark the epochs when $\gamma$-ray flares were detected. Left: overview of the full time range. Right: zoom-in of the region outlined by the black rectangle in the left panel. The two y-axes give the BH separation in projected angular units (left, mas) and in deprojected distance in units of $10^{6}R_g$ (right), where $R_g$ is the gravitational radius of the central black hole.\label{fig:model_vs_time}}
\end{figure*}

To quantify the kinematic behavior described above, we modeled the positions of the jet knot components identified in \autoref{fig:EATING_jets} and \autoref{fig:MOJAVE_jets} using Gaussian components. The resulting trajectories, shown in \autoref{fig:model_vs_time}, confirm the collision and subsequent downstream propagation discussed above.
For the positions of these Gaussian models, refer also to \autoref{fig:image_with_model}.

\begin{table}[!htbp]
  \centering
  \caption{The core shift values of 3C 111 are shown at each frequency. These values represent the distance from the SMBH.}
  \begin{tabular}{lc}
    \toprule
    Frequency (GHz) & Distance from the SMBH (\muas) \\
    \midrule
    43 & $118\pm20$ \\
    22 & $215\pm20$ \\
    15 & $303\pm20$ \\
    \bottomrule
  \end{tabular}
  \label{tab:core_shift}
\end{table}
We corrected the offset in core-position caused by opacity effects using \autoref{tab:core_shift}. 
We assumed that the absolute core shift values follow a power-law dependence on frequency, $\Delta s_{(\nu)} \propto \nu^{-1/\kappa}$ \citep{Porcas2009}, and obtained $\kappa = 1.12 \pm 1.31$. For further details, see \autoref{met:Hough_transform}.
Individual knots can be cross-identified smoothly between epochs, although their apparent speeds vary slightly with distance from the core. 
The linearly propagating and brightening K2 knot, marked by squares in \autoref{fig:model_vs_time}, is a newly identified feature that emerges during our monitoring period.
This feature progressively collides and blends with the K4 knot marked by circles, consistent with a collision between a faster, later-emitted knot and a slower, earlier knot in an internal shock scenario. We therefore interpret this behavior as direct evidence that we are observing an internal shock.
In addition, two knots (labeled K1 and K3 in \autoref{fig:model_vs_time}) appear and propagate in association with the K2 knot. 
The K1, K2, and K3 knots emerge in rapid succession within the same time period, become resolved at all frequencies, and propagate while maintaining similar relative separations and exhibiting comparable trends in apparent speed. Although K1 and K3 knots are faint and not clearly visible in some imaging epochs, model fitting consistently resolves them with sufficient reliability, making it unlikely that they are spurious artifacts.
The ``head section'' before $\sim$2022 appears stable yet highly complex. This complexity hampers robust cross-identification of K1, K2, and K3 knots across epochs, making it difficult to trace the origin of these knots. 
Their inferred knot Lorentz factors are listed in \autoref{tab:gamma_component}, computed assuming a viewing angle of $16.3 \pm 2.3^\circ$ for the 3C~111 jet \citep{Jorstad17}. 

\begin{table}[!htbp]
  \centering
  \caption{Intrinsic Lorentz factor $\Gamma$ and apparent velocity $\beta_\text{app}$ of knot components.}
  \begin{tabular}{lcc}
    \toprule
    Component & $\Gamma$ & $\beta_\text{app}$ \\
    \midrule
    K1 knot & $3.64\pm0.18$ & $3.50\pm0.19$ \\
    K2 knot & $4.28\pm0.16$ & $4.09\pm0.07$ \\
    K3 knot & $4.74\pm0.29$ & $4.44\pm0.09$ \\
    K4 knot & $1.57\pm0.10$ & $0.83\pm0.11$ \\
    \bottomrule
  \end{tabular}
  \label{tab:gamma_component}
\end{table}

\bigskip

\subsection{Flux Evolution of the Knot Components}

\begin{figure}[!htbp]
  \centering
  \includegraphics[width=\linewidth]{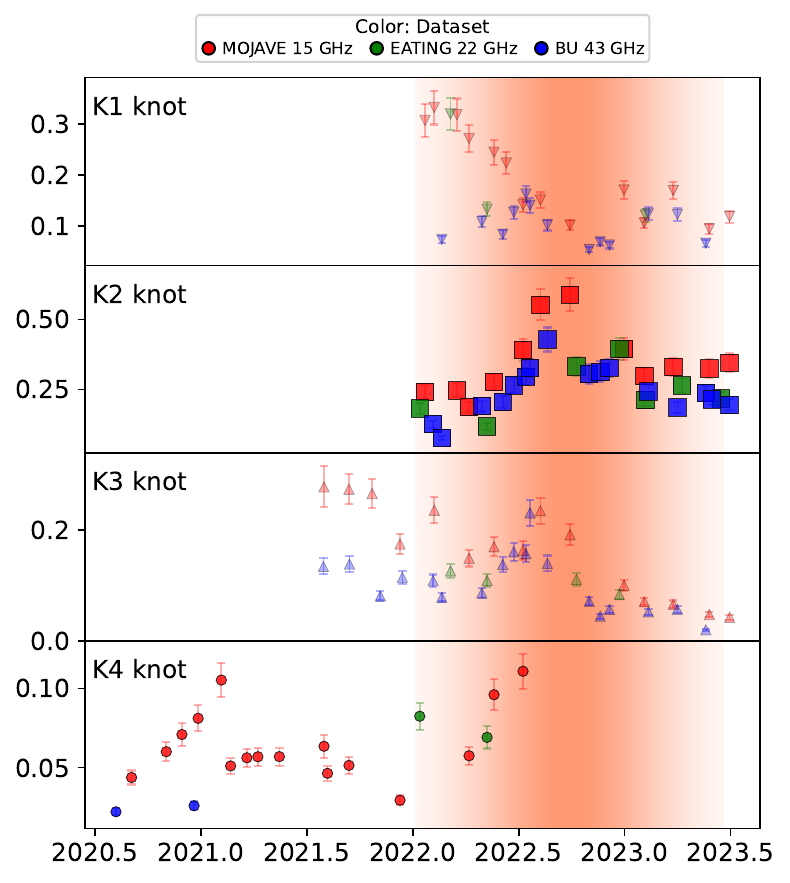}
  \caption{Light curves of knot components in a selected time period. The horizontal and vertical axes show time and peak flux density, respectively. The four panels and the symbol shapes of the data points represent the K1, K2, K3, and K4 knot components, respectively. Marker color indicates the dataset (MOJAVE 15 GHz, EATING VLBI 22 GHz, or BU 43 GHz). The faint red band indicates the full width at half maximum (FWHM) of the Gaussian fit to the peak of the 15 GHz light curve of the K2 knot component. The error bars combine the random uncertainties estimated from the MCMC analysis with a separate 10\% per-epoch flux-density-scale uncertainty; this is distinct from the data-set-specific fractional error floor used in the visibility likelihood.\label{fig:light_curve}}
\end{figure}

\autoref{fig:light_curve} shows the light curves of the knots highlighted in \autoref{fig:model_vs_time}. The K2 knot brightens sharply, reaching its peak flux in mid-2022, and then fades from late 2022 through the end of our monitoring.
This brightening event coincides with the timing of the collision-like feature seen in \autoref{fig:EATING_jets}, \autoref{fig:MOJAVE_jets}, and \autoref{fig:model_vs_time}, supporting the occurrence of an actual internal shock collision. 
In contrast, the K3 knot may suggest only a modest brightening around the epoch when they approach the K2 component, but the enhancement is much weaker than that associated with the K2 knot. 
This implies that the brightness of the propagating bright feature seen in \autoref{fig:EATING_jets} and \autoref{fig:MOJAVE_jets} is primarily dominated by the K2 knot.
Although the K1 and K3 knots, like the K2 knot, geometrically collide with the K4 knot, they do not exhibit pronounced brightening comparable to that of the K2 knot. This difference may arise from variations in magnetization and/or number density among individual knot components\citep{Mimica12, Spada01}.
Focusing on the K2 knot component in \autoref{fig:light_curve}, the flux reaches its peak at approximately the same epoch across all observing frequencies. However, in the decay phase, the 43 GHz light curve appears to decline earlier than the 15 GHz light curve. This behavior is qualitatively
consistent with the theoretical expectation that higher-energy non-thermal electrons cool on shorter timescales \citep{Sari98}. However, it is difficult to discuss possible frequency-dependent differences during the rising phase. 
{
In addition, the K4 knot in \autoref{fig:light_curve} shows an increase in flux while K2 knot is still identified as a separate component but is approaching K4 knot in \autoref{fig:model_vs_time}. We interpret this early brightening not as a consequence of the final blending epoch itself, but as evidence that the interaction between the two components had already begun before they became indistinguishable in the model fitting. As their projected separation decreased, gradual compression of the emitting plasma associated with K2 and K4 knots could have enhanced the flux of both components, supporting a causal connection between the brightening and the interaction of the two knots.
Such continuous brightening driven by successive interactions was also observed in the kpc-scale internal shock collision reported by \citet{Meyer2015}.
}

\bigskip

\subsection{Polarization Signatures of the Shock}

\begin{figure*}[!tp]
\centering
\includegraphics[width=0.82\textwidth]{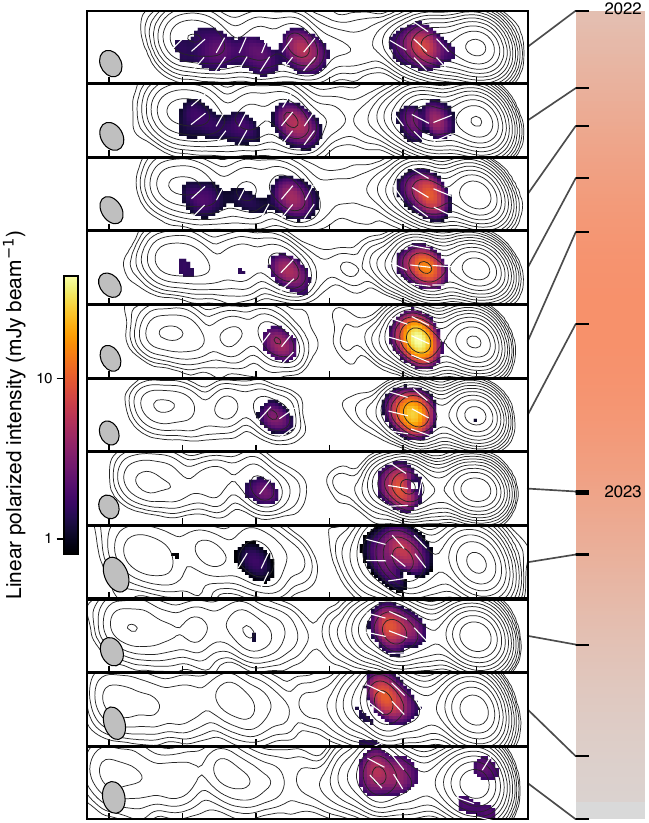}
\caption{Archival MOJAVE CLEAN polarization maps at 15~GHz for 11 epochs from 2022 April to 2023 July. Stokes~I contours begin at $3\sigma_I$ and increase by factors of two. Color represents the observed linear polarized intensity, $P_{\rm raw}=(Q^2+U^2)^{1/2}$, in mJy~beam$^{-1}$ on a logarithmic scale, and the white ticks show the electric vector position angle (EVPA). The polarization color and EVPA ticks are displayed only where $I\geq5\sigma_I$ and $P_{\rm raw}\geq5\sigma_P$; no additional lower cutoff in fractional linear polarization is applied. The CLEAN restoring beam for each epoch is shown in the lower-left corner of its panel, and the horizontal tick marks are separated by 2~mas. The images are aligned on the core and rotated $25^\circ$ counterclockwise for display. The gray band to the right gives the observing timeline, and the faint red band indicates the full width at half maximum of the Gaussian fit to the MOJAVE K2 light-curve peak shown in \autoref{fig:light_curve}.\label{fig:pol_montage}}
\end{figure*}

Linearly polarized images provide another powerful diagnostic of shock activity. Because the cross-hand visibilities have low signal-to-noise ratios, we conservatively based the polarization analysis on the publicly available MOJAVE archival CLEAN Stokes~$I$, $Q$, and $U$ images. \autoref{fig:pol_montage} shows these images at 11 epochs contemporaneous with the putative internal shock event.
In the archival CLEAN images, the fractional linear polarization measured in the fixed K2-centered aperture increases from approximately $0.9\%$ in 2022 April--May to a maximum of approximately $6.6\%$ in 2022 August. Because K2 is not fully resolved from the neighboring knots during the polarization maximum, this value characterizes the K2-centered interaction region rather than K2 alone.
We also applied the fixed-aperture measurement to the Gaussian-model positions of the other knot components. Among aperture-level polarized-intensity detections above $5\sigma_P$, we obtained fractional-linear-polarization ranges of $0.34$--$3.16\%$ for K1, $0.41$--$5.84\%$ for K3, and $1.56$--$5.92\%$ for K4, which was identified in only three of the polarization epochs. These values do not represent independent component measurements because of beam blending. In particular, K2 and K3 are unresolved in the beam when the K3-centered aperture reaches $5.74\%$ and $5.84\%$ in 2022 August and September, respectively; K4 overlaps K3 within both the aperture and the beam in all three available epochs; and K1 is unresolved from K2 in many epochs (their separation in 2022 September is $0.522$~mas, compared with a beam width of $0.534$~mas along their separation). In the two epochs in which K3 is geometrically separated from the neighboring knots, 2023 May and July, its aperture-level polarized-intensity signal-to-noise ratios are only $4.04$ and $2.74$, respectively, and therefore do not meet the $5\sigma_P$ detection criterion.

\bigskip

\begin{figure}[!htbp]
\centering
\includegraphics[width=0.95\linewidth]{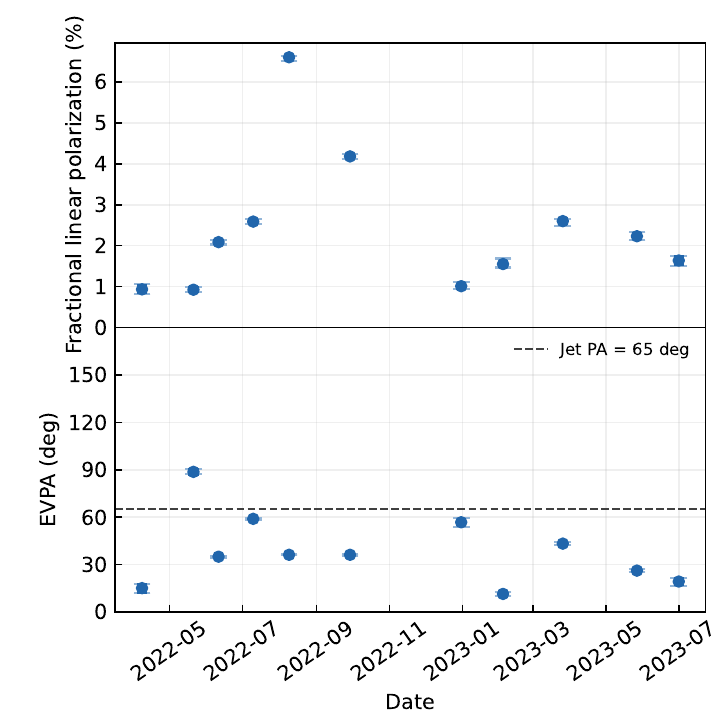}
\caption{Polarization measured in a fixed 0.25-mas-radius aperture centered on the aligned K2 position in the archival MOJAVE CLEAN maps. Top: fractional linear polarization. Bottom: electric vector position angle (EVPA). The Stokes~$I$, $Q$, and $U$ values are averaged over the full fixed aperture without a polarized-intensity-based pixel selection. The narrow dark error bars show the 68\% intervals from the published map noise, while the wider pale error bars additionally include the uncertainties in the Gaussian-model K2 position and the alignment between the Gaussian model and the CLEAN Stokes~I map. The horizontal dashed line indicates the jet position angle, $65^\circ$. No Faraday-rotation correction has been applied.}
\label{fig:pol_curve}
\end{figure}

The temporary increase in fractional linear polarization in the K2-centered interaction region is qualitatively consistent with enhanced magnetic-field ordering or compression associated with a shock \citep{Marscher2008}. We use only this increase as the polarization signature of the event and do not use the EVPA evolution to infer the detailed shock geometry.
Finally, we note that no Faraday rotation correction was applied to the images.
A recent multi-frequency polarimetric study\citep{Bartolini2025} detected a high rotation measure of a few thousand~rad\,m$^{-2}$ in these regions, which may affect the observed EVPA.
Furthermore, relativistic aberration can systematically rotate the EVPA \citep{Park26}, and this too remains uncorrected.
Consequently, any discussion of the absolute EVPA is inherently limited. Therefore, we do not pursue detailed modeling here, but future theoretical work may be important to explore this scenario further.
\bigskip

\section{Discussion}
\label{sec:discussion}

{

\subsection{Lack of Post-Collision Deceleration}

The K2 knot retains a high Lorentz factor ($\Gamma \sim 4.3$) even after its brightness peak in September 2022. In contrast, 
the two-mass collision model used in discussions of $\gamma$-ray emission in blazars, in which the event is treated as an inelastic collision between K2 and K4 knots, generally predicts that the K2 knot decelerates after the collision \citep{Spada01}. For example, if we assume a mass ratio of 10 between the K2 and K4 knots, a K2 knot with $\Gamma \sim 4.3$ would decelerate to $\Gamma \sim 3.8$ after the collision.
To avoid significant deceleration even after the collision, some form of energy injection is required. This energy may originate either from magnetic energy carried by the jet \citep{Blandford2019} or from thermal energy generated by the collision \citep{Ricci24}.
}

{\subsection{Why this is not a standing shock}

One might interpret the interaction between the K2 and K4 knots as a standing shock.
}
However, the K2 knot is not stationary: it propagates with an inferred relativistic speed of $\Gamma \sim 4$--$5$. If this feature were instead a standing shock, an additional mechanism would be required to anchor it at a fixed location, while also producing the observed K2 knot. Moreover, after the interaction, we do not observe the emergence of a new quasi-stationary knot at the position expected for the original K4 knot.
{
If this feature were instead part of a sequence of intermittent diamond shocks, similar structures would be expected to appear persistently along the jet; however, no such signature is evident in the present observations.
}
Taken together, these considerations favor an interpretation of the K2 knot corresponding to an internal shock rather than a standing shock.
{
However, the component shown in Figure~5 of \citet{Beuchert2018} that reached this region exhibited no increase in total intensity, and no geometrical collision with another knot component was identified.
}
There is precedent for this kind of phenomenon, although the previously reported case occurred on kiloparsec scales \citep{Meyer2015} rather than parsec scales.
We summarize our interpretation in \autoref{fig:cartoon_internal_shock}.
Our results support the coexistence of a quasi-stationary standing shock at sub-pc scales and a transient internal shock propagating at larger distances.

\bigskip

\subsection{Origin of the 2021 Gamma-Ray Flare}

\begin{figure}[!htbp]
\centering
\includegraphics[width=0.9\linewidth]{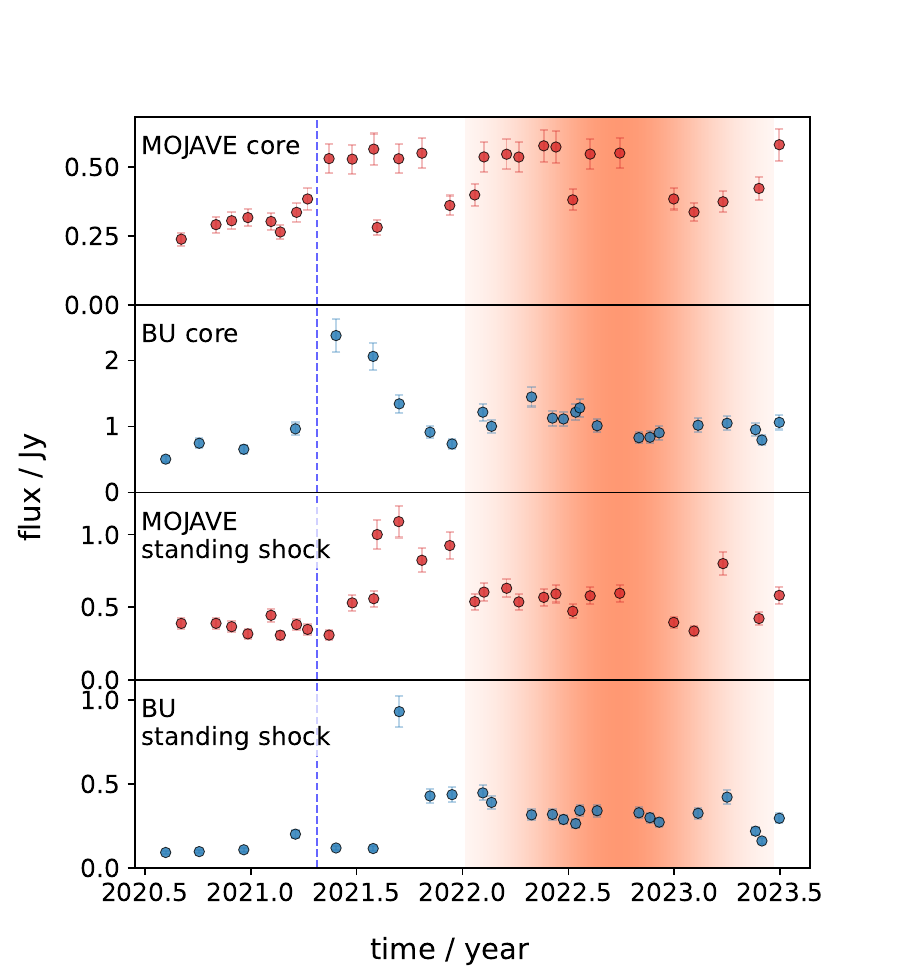}
\caption{
Gaussian-modeled light curves of the core and standing shock. The four panels correspond to the core and standing-shock components in the MOJAVE and BU data sets. The horizontal axis shows time, and the vertical axis shows variations in the flux of each component. The blue vertical dotted line marks the prominent $\gamma$-ray flare in April 2021, while the light red shaded band indicates the period when the K2 knot brightened due to an internal shock. The error bars combine the random uncertainties estimated from the MCMC analysis with a separate 10\% per-epoch flux-density-scale uncertainty; this is distinct from the data-set-specific fractional error floor used in the visibility likelihood.
}
\label{fig:light_curve_core}
\end{figure}

In radio jets—particularly those of blazars—the occurrence of many $\gamma$-ray flares has been reported to correlate with the epoch at which a newly formed knot passes through a radio core \citep{Marscher2008, Lico22, Casadio19, Jorstad16, Marscher10, Kramarenko22}.
Similar high-energy activity associated with interactions between moving knots and standing shock components has also been discussed in other systems, for example HST-1 in M87 \citep{Giroletti12}, S5 0716+714 \citep{MAGIC18}, and PKS 1510$-$089 \citep{HESS21}.
As mentioned above, earlier observations established the presence of a standing shock in 3C~111 at sub-parsec projected distances from the central engine \citep{Jorstad16, Schulz2020}.
The light curve of the core (see \autoref{fig:light_curve_core}) shows that the core brightened simultaneously with the $\gamma$-ray flare in 2021, while the standing shock component brightened with a slight delay. This behavior is consistent with a scenario in which the K2 knot first interacted with the core and subsequently with the standing shock located farther downstream.

Therefore, the 2021 April GeV $\gamma$-ray flare is more likely to be associated with the interaction of K2 with the core than with its subsequent interaction with the standing shock.
The ejection time of the K2 knot from the immediate vicinity of the SMBH, estimated by back-extrapolating from \autoref{fig:model_vs_time}, is April 20, 2021 $\pm$ 7.5 days (2021.30 $\pm$ 0.02 year), which is also consistent with the above scenario.

\bigskip

\subsection{High-Energy Implications of a Parsec-Scale Internal Shock}

\begin{figure*}[!tp]
\centering
\includegraphics[width=0.92\textwidth]{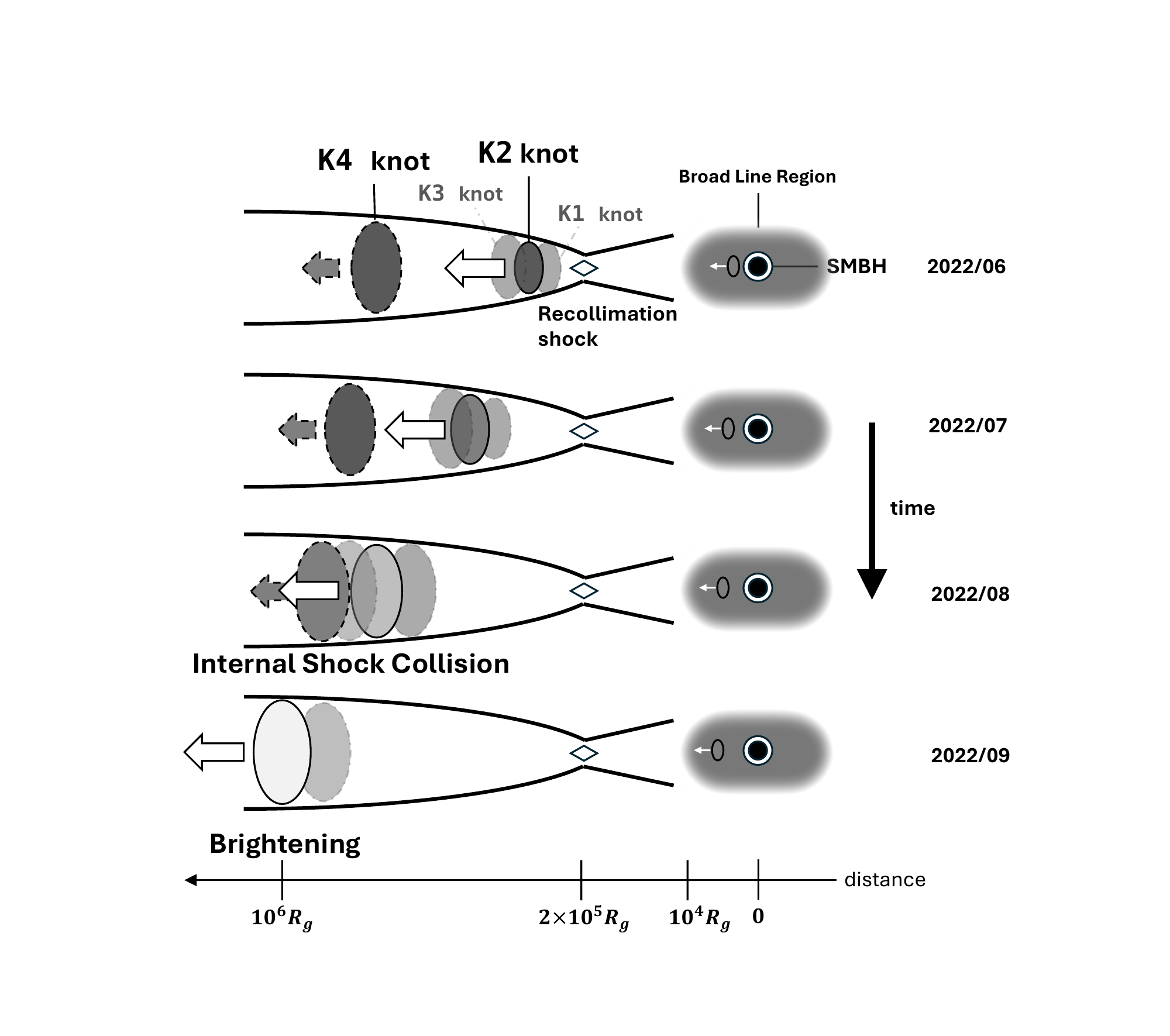}
\caption{Schematic cartoon illustrating the evolution of an internal shock in the jet of 3C 111. The figure consists of four jet panels corresponding to four epochs, with time progressing from top to bottom. The rightmost column lists the observing date for each panel. In each panel, the K1, K2, K3, and K4 knots propagate downstream of the recollimation shock; their interaction eventually produces an internal shock, leading to brightening. Each jet panel also shows the supermassive black hole, a broad-line region, and a recollimation shock, together with a jet width that systematically varies with distance from the black hole. The recollimation shock is also referred to as a standing shock or a reconfinement shock. The bottom axis shows the deprojected distance from the black hole to each point in units of gravitational radii($R_g$). The broad-line region (BLR) is rich in seed photons, and $\gamma$-ray brightening is expected if an internal shock occurs within this region.\label{fig:cartoon_internal_shock}}
\end{figure*}

A key feature of the internal shock identified in this study is that its location can be directly determined with VLBI. 
Because the distance from the central engine is known, this provides a valuable test of the internal shock scenario.
The test focuses on whether high-energy emission brightens simultaneously with the radio flare associated with the internal shock. Such emission is most naturally produced by inverse-Compton scattering of soft seed photons near the internal shock site, and thus its presence depends on the energy density of the soft-photon field. However, the internal shock occurs at a deprojected distance of about 10 pc from the central engine, where the energy densities of photons from the broad-line region (BLR) and the accretion disk are expected to be low. Inverse-Compton scattering is therefore inefficient, and no accompanying high-energy flare is expected.
Consistently, no significant enhancement is seen in the Fermi/LAT light curve, supporting the internal shock scenario.
As shown in \autoref{fig:cartoon_internal_shock}, the BLR is located very close to the black hole—about two orders of magnitude interior to the internal shock we detect and also inside the 43 GHz core \citep{Bartolini2025}. An internal shock occurring within the BLR is therefore expected to be accompanied by a $\gamma$-ray flare. Consequently, among the $\gamma$-ray flares from radio jets, those that are not associated with passage of a knot through a standing shock may instead be explained by internal shocks.

\section{Summary and Conclusion}
\label{sec:conclusion}

We have presented a 64-epoch, three-frequency VLBI study of the nearby radio galaxy 3C~111 spanning 2020--2023, combining 22~GHz EATING VLBI data with 15~GHz MOJAVE and 43~GHz BU archival data. Together with the archival MOJAVE CLEAN polarization images at 15~GHz, these observations allow us to track the morphology, kinematics, and flux evolution of knot components and the polarization behavior of the interaction region through the development of a parsec-scale internal shock.

Our main findings are summarized as follows.
\begin{enumerate}
  \item The multi-frequency image sequence reveals a direct internal shock generated as a faster, trailing knot (K2) catches up with a slower, leading knot (K4). The internal shock begins around April 2022 and is followed by marked brightening of K2 at a projected distance of 2--3 pc from the core, corresponding to a deprojected distance of about 10 pc from the central engine.

  \item Gaussian component modeling confirms that K2 is a relativistically moving knot rather than a quasi-stationary feature. In contrast, the standing shock identified in previous studies remains consistent with a recollimation-shock interpretation. The observed behavior therefore favors a scenario in which a transient internal shock coexists with a quasi-stationary standing shock in the jet of 3C~111.

  \item The flux evolution provides independent support for the internal shock interpretation. K2 shows the clearest brightening among the tracked knots, and its flux peaks at approximately the same epoch across all three observing frequencies. The subsequent earlier decline at 43 GHz than at 15 GHz is qualitatively consistent with faster cooling of higher-energy electrons.

  \item The archival MOJAVE CLEAN polarization images show a temporary increase in fractional linear polarization from approximately $0.9\%$ to a maximum of approximately $6.6\%$ in the K2-centered interaction region. Because K2 is not fully resolved from the neighboring knots at the polarization maximum, the peak value does not represent K2 alone. We use this increase as an additional polarization signature of the event without inferring a point-by-point correspondence with the Stokes~I flux or a detailed interpretation of the EVPA evolution.

  \item The 2021 April GeV $\gamma$-ray flare is more naturally associated with activity near the core than with the later parsec-scale internal shock. The inferred ejection time of K2 from the immediate vicinity of the SMBH is consistent with this timing, and the core light curve brightens contemporaneously with the flare, whereas the standing shock brightens later.

  \item No significant high-energy enhancement is observed during the parsec-scale internal shock itself. This is consistent with the internal shock taking place far downstream of the broad-line region and accretion-disk photon fields, where inverse-Compton scattering is expected to be inefficient. Our results therefore support a picture in which internal shocks can occur on parsec scales without producing a contemporaneous $\gamma$-ray flare, while internal shocks occurring much closer to the black hole could still contribute to high-energy activity.
\end{enumerate}

Taken together, these results provide spatially resolved observational evidence that an internal shock can be identified directly in an AGN jet. In 3C~111, the jet appears to host both a standing shock near the core and a downstream internal shock produced by the interaction of moving knots. This hybrid picture offers a natural framework for interpreting the diversity of radio and high-energy variability seen in relativistic jets.

Future higher-resolution, higher-cadence multi-frequency VLBI monitoring---particularly with polarization information closer to the core---will be crucial for determining how often such internal shock events occur, how they are triggered, and under what conditions they generate observable high-energy flares.

%
\appendix

\section{Hough transform}
\label{app:hough}
\setcounter{figure}{0}
\renewcommand{\thefigure}{\Alph{section}.\arabic{figure}}
\renewcommand{\theHfigure}{\Alph{section}.\arabic{figure}}

\begin{figure}[t]
  \centering
  \includegraphics[width=\linewidth]{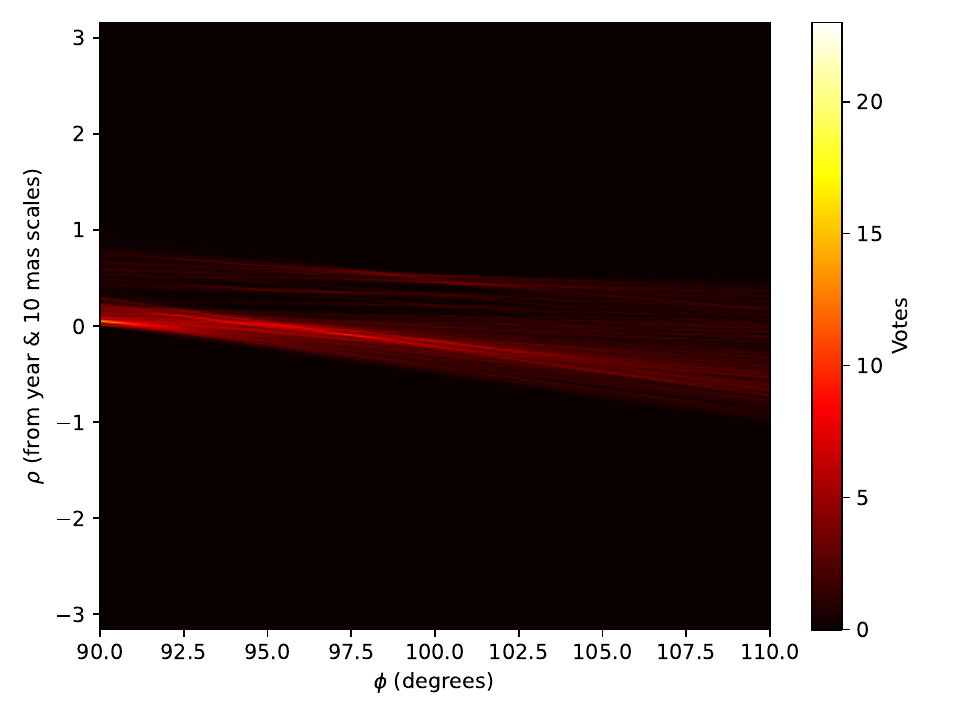}
  \caption{Accumulator heatmap for the resulting core-shift value. The color indicates the number of votes. The horizontal and vertical axes indicate the parameters $\phi$ and $\rho$, respectively. For details of the Hough transform method, see \autoref{met:Hough_transform}.}
  \label{fig:accumulator}
\end{figure}

{
\autoref{fig:accumulator} displays the accumulator of resulting core shift value shown in \autoref{tab:core_shift}. 
The explored range of $\phi$ corresponds to a typical apparent velocity of $\sim 8c$ for this source \citep{Jorstad17}.
The region with $\rho \le 0$ corresponds to lines that identify components ejected from the core after 2021.5, suggesting that this data set contains more components ejected after 2021.5 than before 2021.5. The local peaks smoothly distributed from $90^\circ$ to $110^\circ$ indicate that, at each angle, lines connecting specific pairs of components are systematically identified compared with random combinations of parameters. The strongest validation of the resulting core-shift value is that \autoref{fig:model_vs_time}, to which this result is applied, connects smoothly across frequencies and appears consistent with the physical behavior of the jet.
}

\bibliographystyle{aasjournalv7}
\bibliography{references}

\section*{Acknowledgments}

K.A. acknowledges financial support from the National Science Foundation (AST-2034306, AST-2107681, AST-2132700, and AST-2535855) and the Gordon and Betty Moore Foundation (GBMF-12987). K.H. acknowledges support from MEXT/JSPS KAKENHI (Grant Numbers 25H00660, 22H00157, and 21H04488), the Mitsubishi Foundation (Grant Number 202310034), and the Daiko Foundation (Grant Number J0SE807004). L.C. acknowledges support from the Tianshan Talent Training Program ( Grant Number 2023TSYCCX0099). This work was supported in part by a University Research Support Grant from the National Astronomical Observatory of Japan and by the Grant-in-Aid for Outstanding Research Group Support Program at Nagoya City University (Grant Number 2530002). K.K. acknowledges support from JST SPRING, Grant Number JPMJSP2131. This research has made use of data from the MOJAVE database that is maintained by the MOJAVE team. This research has also made use of publicly available Fermi-LAT data from the LAT Light Curve Repository. This research has also made use of VLBA data from the VLBA-BU Blazar Monitoring Program (BEAM-ME and VLBA-BU-BLAZAR), funded by NASA through the Fermi Guest Investigator Program. The VLBA is an instrument of the National Radio Astronomy Observatory, a facility of the National Science Foundation operated by Associated Universities, Inc. Badary, Svetloe, and Zelenchukskaya radio telescopes are
operated by the Scientific Equipment Sharing Center of the Quasar VLBI
Network.

\end{document}